\documentclass[12pt,preprint]{aastex}

\usepackage{color}

\slugcomment{

Submitted to ApJ 17 June 2004.\\
Manuscript pages : \pageref{'last page'}, Figures: \ref{fig:observability}\\

}
\shorttitle{Light curves for extrasolar planets}
\shortauthors{Dyudina et al.}

\def\mm{$\mu$m }
  
\def\tms{$\times$}  
\def\ie{{ i.e.}}  
\def\eg{{ e.g.}}  
  
\def\deg{$^{\circ}$}
\def\nh3{ $\mbox{NH}_3$}
\def\ch4{ $\mbox{CH}_4$}
\def\h2o{ $\mbox{H}_2\mbox{O}$}
\def\c2h2{ $\mbox{C}_2\mbox{H}_2$}

\def\cora#1{\textbf{\textcolor{black}{\rm#1}}}   
\def\corb#1{\textbf{\textcolor{black}{\rm#1}}}   
\def\corc#1{\textbf{\textcolor{black}{\rm#1}}}

\begin{document}
\bibliographystyle{apj}
 
\title{Phase light curves for extrasolar Jupiters and Saturns}
\author{Ulyana A. Dyudina\altaffilmark{1,2}, Penny D. Sackett\altaffilmark{2}, Daniel D. R. Bayliss\altaffilmark{3}}
\affil{Research School of Astronomy and Astrophysics, Mount Stromlo Observatory, Australian National University, Cotter Road, Weston, ACT, 2611, Australia}

\author{Sara Seager}
\affil{DTM, Carnegie Institute at Washington, DC , USA}

\author{Carolyn C. Porco}
\affil{CICLOPS/Space Science Institute, Boulder, CO, USA}


\author{Henry B. Throop and Luke Dones}
\affil{Southwest Research Institute,Boulder, CO, USA}

\altaffiltext{1}{NASA/Goddard Institute for Space Studies,
2880 Broadway, New York, NY, 10025, USA}
\altaffiltext{2}{Planetary Science Institute, Australian National University, ACT, 0200, Australia}
\altaffiltext{3}{Victoria University of Wellington, New Zealand }

\begin{abstract}
We predict how a remote observer would see the brightness variations of giant planets similar to those in our Solar System as they orbit their central stars.
We model the geometry of Jupiter, Saturn and Saturn's rings for varying orbital and viewing parameters. Scattering properties for the planets and rings at wavelenghts 0.6-0.7 microns are assumed to follow these observed by Pioneer and Voyager spacecraft, namely, planets are forward scattering and rings are backward scattering. Images of the planet with or without rings are simulated and used to calculate the disk-averaged luminosity varying along the orbit, that is, a light curve is generated. We find that the different scattering properties of Jupiter and Saturn (without rings) make a substantial difference in the shape of their light curves. Saturn-size rings increase the apparent luminosity of the planet by a factor of 2-3 for a wide range of geometries, an effect that could be confused with a larger planet size. Rings produce asymmetric light curves that are distinct from the light curve of the planet without rings, which could resolve this confusion. If radial velocity data are available for the planet, the effect of the ring on the light curve can be distinguished from effects due to orbital eccentricity. Non-ringed planets on eccentric orbits produce light curves with maxima shifted relative to the position of the maximum planet's  phase. Given radial velocity data, the amount of the shift restricts the planet's unknown orbital inclination and therefore its mass. Combination of radial velocity data and a light curve for a non-ringed planet on an eccentric orbit can also be used to constrain the surface scattering properties of the planet and thus describe the clouds covering the planet. To summarize our results for the detectability of exoplanets in reflected light, we present a chart of light curve amplitudes of non-ringed planets for different eccentricities, inclinations, and the viewing azimuthal angles of the observer. 

\end{abstract}


\section{Introduction}

Modern space-based telescopes and instrumentation are now approaching the
precision at which reflected light from extrasolar planets can be
detected directly \citep{jenkins03,walker03,green03,hatzes03}. 
Since 1995, meanwhile, more than 100 extrasolar planets (or exoplanets) have been detected indirectly by measuring the reflex motion of their parent star along the line of site (radial velocity or Doppler method).  
One of these radial velocity planets, Gl876b, has been confirmed by measuring the parent star motion on the sky (astrometry) \citep{benedict02}, and a second, HD209458b, by measuring the change in parent star brightness as the planet executes a transit of its host (transit photometry) \citep{charbonneau00,henry00,udalski02a,udalski02b,udalski03}. 
Three exoplanets have now been detected by transit photometry \citep{konacki03,konacki04,bouchy04},
and then confirmed with Doppler measurements.  
It is expected that in the next decade, as photometric techniques are improved both from the ground and in space, direct detection of the reflected light from exoplanets will not only
prove useful in expanding the number of exoplanets known, but also in detailing their characteristics.\footnote{ Reviews and references on
extrasolar planets and detection techniques can be found at {\it
http://www.obspm.fr/encycl/encycl.html}}

Reflected light from an exoplanet can be detected in two ways.  First, with sufficient spatial resolution, direct imaging can resolve the planet and the star in space, and the projection of its orbit traced as a function of time simultaneously with the measurement of the planet's phases in reflected light.  
To be detected, the planets must be at large enough orbital distances to be easily resolved from their parent star, and yet close enough that the reflected brightness, which
decreases as inverse square of the orbital distance, is large enough to be observed against the background in a reasonable amount of time.
Even with coronagraphic or nulling techniques to help block the light at the position of the star itself, success will require optical point spread functions with very low level scattering wings to obtain sufficient reduction of parent starlight at the position of the
planet.  
Consequently, the first extrasolar planets to be detected directly in
this way are likely to be giants orbiting relatively nearby stars (tens of pc) at intermediate semi-major axes (1-5 AU) 
\citep{dekany04,lardiere03,codona04,trauger03,krist03,clampin01}, 
where the contrast between planet and scattered starlight is high enough and yet the total reflected light still measurable. 
Since direct imaging 
\corb{can} 
yield both the orbit and the luminosity of the planet simultaneously, 
it will be a robust detection method that will add greatly to our knowledge of planetary structure and evolution; 
Unfortunately direct exoplanet imaging from either the ground or space will not likely be available for a decade.  
In this paper, therefore, the applicability of our results to direct imaging is not described in detail, although our light curves can be used for planning future observations aimed at directly detecting spatially-resolved planets in reflected light.

The second method of detecting reflected light is precise, integrated
photometry.  
This technique searches for temporal variations in the combined parent star and
exoplanet (reflected) light curve as the planet changes phases during
the orbit in the same way the Moon changes phase, producing a``phase
light curve.''  The vast majority of the light comes from the star
itself, but if the periodic variations can be extracted from the total
light curve, the planetary phase as a function of time can be deduced.
Since the light from the parent star need not be blocked, if the signal 
to noise is large enough, 
planets much closer to their parent star can be detected then with spatially resolved imaging.  
Futhermore, in principle, more distant planetary systems and planets at smaller physical 
orbital radii can be studied in this way since the angular size of the point spread function does not present a limitation.  

Ground-based based observations of known short-period exoplanets 
have used Doppler tomographic signal-analysis techniques to search 
for reflected light signatures in high-resolution spectroscopy.  
A combined, high signal-to-noise spectrum of the star is shifted 
and subtracted from individual spectra taken at different times (\ie,  
different phases of the exoplanet's illumination).  Since the 
planet contributes a different amount of reflected light to 
the combined spectrum as a function of orbital phase, one expects 
a residual in the subtracted spectra at a (exoplanet) velocity position 
that varies in sinusoidally with phase.  In every case attempted 
to date, no signature of reflected light has been detected 
\citep{charbonneau99, ccameron02, 
leigh03a, leigh03b}.  
However, because the orbital phase is known from radial velocity measurements, the 
lack of detection constrains the grey geometric albedo $p$
of the planet 
\corb{for an assumed phase curve}, orbital inclination, and planetary radius.  
For the assumed parameters of $\tau$\ Bootis\ b, HD\ 75289b, and the innermost planet of $\upsilon$\ And, observations non-detections seem to imply that 
$p< 0.4$ \citep{ccameron02,leigh03a, leigh03b}.

Current ground-based observations for reflected starlight from 
exoplanets are able to reach a precision of $\sim$$10^{-5}$ \citep{leigh03a,leigh03c}, 
if the planetary reflection spectrum is fairly well known.  
Current and near-future space-based telescopes are
capable of detecting variations of the order $10^{-4}-10^{-6}$ of the
total star light \citep{jenkins03, green03}, at the limit of what
might be expected from the class of "hot Jupiter" planets, \ie, gas
giants orbiting within 0.1 AU of their solar-type parent stars \citep{seager00}.  
With the advent of 30m-100m telescopes in the next 10-15 years, one might expect these limits to improve by factors of 10 to 100.

The shape of the phase light curve depends on many parameters: 
the planet's orbit, its geometry relative to the observer, \corc{the} planet's oblateness, the scattering properties of the planet's surface, and the presence and geometry of rings.
The amplitude of the light curve scales in a simple manner 
with planet's size and its orbital distance. 
When planning observations, it is important to understand the variety of shapes and amplitudes of these phase light curves and the conditions under which they will be generated: this is the purpose of this paper.  
In particular, the effects of planet size, oblateness, rings, and ``surface'' scattering properties, which convey information about atmospheric structure and reflecting clouds, will be superimposed in 
\corc{observed} light curves and may be indistinguishable from one other, causing confusion or degeneracies in interpretation.
We discuss these degeneracies and possible ways to eliminate them through detailed light curve measurements or alternative observations.

The great interest in the potential of exoplanet detection 
\corb{through either reflected light, reflected phase light curves, or transits}
is demonstrated by a number of works published in the last few years. 
\cite{seager00} modeled the atmospheric and cloud composition and 
simulated the light curves for close-in giant planets with various cloud coverage.
\corb{\cite{sudarsky03} modeled clouds and spectra for extrasolar planets at different distances from the star.}
\cite{barnes03} modeled transit light curves for a planet with rings. 
\cite{arnold04} simulated light curves for planets with rings of different sizes, 
assuming planets with isotropically-scattering (Lambertian) surfaces and rings with isotropically scattering particles, and provide a discussion of the possibility of ring presence at different stages of planetary system evolution.
 
Our model is the first to use the {\it observed anisotropic scattering\/} of Jupiter, Saturn, and Saturn's rings.  
We find that anisotropic scattering yields light curves that are substantially
different from those assuming Lambertian planets and rings, and thus
is more likely to give an accurate description for extrasolar planets
similar to Jupiter and Saturn.  We do not model the spectral light
curve dependence; our model produces light curves for averaged red visible light ($\sim$0.6-0.7\mm). 

In Section \ref{sec:model} we describe our model. 
In Section \ref{sec:results} we present our light curves for non-ringed exo-Saturn 
and exo-Jupiter planets on an edge-on circular orbit, 
\corc{for} variously oriented oblate exo-Saturns, for a ringed planet (with Saturn-like scattering properities) at variously inclined circular orbits, and 
for both a non-ringed exo-Saturn and an exo-Jupiter on eccentric orbits. 
In Section \ref{sec:discussion}, we discuss the uncertainties of our model and the detectability of the light curves by modern and future instruments.
Conclusions are presented in Section \ref{sec:conclusions}.

\section{Model}
\label{sec:model}

We model the observed red brightness of an exo-Jupiter or 
exo-Saturn by tracing how light rays from the central 
star are reflected by each position on the planet.
We then produce images (maps with resolution of 
16 to 200 pixels across, depending on the acceptable error level) 
of the planet and its rings at different geometries.
The light rays from the star illuminating the planet are assumed to be parallel, consistent with both star and planet being negligible in size compared to the star-planet distance.
The reflected rays collected by observer are assumed to be parallel, consistent with a remote observer. 
The model includes reflection from the planet and rings, and the rings' transmission and shadows, but does not incorporate second order effects such as ring shine on the planet or planet shine on the rings.
\corc{We account for planet's oblateness in some of our simulations using the 10\% oblateness of Saturn as an example}.

\corc{The red scattering properties in our model are the average properties integrated along the planet's spectrum, weighted by the wavelength-dependent transmissivity of the Pioneer red filter, which is non-zero in the range 0.595-0.720\mm.}
Our  notation matches that of most observational papers on Saturn and Jupiter. 
For comparison of our results with those from models of  \cite{arnold04}, we list both our and their 
\corc{parameter} names in Table \ref{tab:def}.
\noindent
\begin{table}[htbp]
\begin{small}
\begin{tabular}{|llll|}
\hline\noindent

This work    & Arnold \&          & Quantity & Units\\
             & Schneider 2004 &          & (if any) \\
\hline\noindent
$A,B$   &        &Coefficients of the Backstorm law                 &\\
$a,a_1$ &        &Semi-major axis of the planet's orbit             &AU\\
$D_P$   &        &Planet-star distance                              &km\\
$E$     &        &Eccentric anomaly (polar angle parameterization)&\\
$e$     &        &Eccentricity                                      &\\
$F$     &$\gamma$&Intensity of a white Lambertian surface\tablenotemark{a}
                                                                    &$Wm^{-2}sr^{-1}$\\
$g_1,g_2,f$ &    &Parameters of Henyey-Greenstein function          &\\
$I$     & $L_p,I_R,I_T$&Intensity (or brightness, or radiance) of the surface&$Wm^{-2}sr^{-1}$\\
        &$L_{ps},L_{pr}$&                                           &       \\
$i$     & $i$    &Inclination of the orbit (0\deg: face-on, 90\deg: edge-on) &degrees\\
$L_P$   &        &Luminosity of the planet\tablenotemark{b}         &$W sr^{-1}$\\
$L_*$   &        &Luminosity of the star\tablenotemark{b}        &$W sr^{-1}$\\
$M$     &        &Mean anomaly (time parameterization)              &\\
$p$     &        &Full-disk albedo\tablenotemark{b}  $L_P/(\pi R_P^2 F)$ &\\
$R$     &        &Star's magnitude in R band                        &\\
$R_P$   & $r_p$  &Equatorial radius of the planet                   &km\\
$r_{\rm pix}$&       &Pixel size                                        &km\\
$V$     &        &Star's magnitude in V band                        &\\
$\alpha$&$\alpha$&Phase angle                                       &degrees\\
$\epsilon$&      &Planet's obliquity                                &degrees\\
$\Theta$&$\phi=\Theta-90$\deg&Orbital angle ($\pm$180\deg: min phase, 0\deg: max phase)&degrees\\
$\mu_0,\mu$&$\mu_0,\mu$&Cosines of the incidence and emission angle&\\
$\omega$&        &Argument of pericentre(90\deg: observed from       &degrees\\
        &        &pericentre, -90\deg: from apocentre)              &       \\
$\omega_r$&      &Observer's azimuth relative to the rings       &degrees\\
\hline     
\end{tabular}
\caption{Parameters used in modeling.}
\label{tab:def}
\end{small}
\tablenotetext{a}{$F\cdot(\pi \rm{~ steradians})$ is the incident stellar flux at the planet's orbital distance (which is also sometimes called $F$, but has $W m^{-2}$ units unlike our intensity $F$ measured per unit solid angle).}
\tablenotetext{b}{The `red' or `blue' optical properties are the convolution of the planet's (or ring's or solar) spectrum with the wavelength-dependent transmissivity of the Pioneer filter (which is nonzero between 0.595-0.720\mm for red filter and 0.390-0.500\mm for blue filter)}
\end{table}


\corc{Figure \ref{fig:angles} demonstrates the geometry fora ringed planet on a circular orbit defined by the angles listed in Table \ref{tab:def}.
\begin{figure}[htbp]
\vspace{-5cm}
\plotone{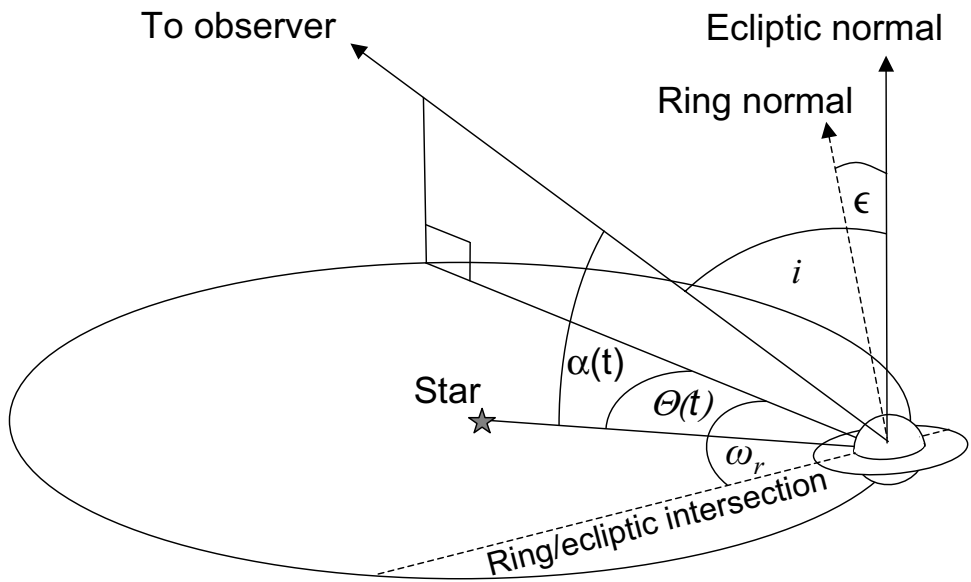}
\vspace{-2in}
\caption{
Angles defining the geometry of a ringed planet on a circular orbit observed from a large distance.
Dashed lines are the normal to the ring plane and the line of intersection of the ecliptic with the ring plane.
The observer's azimuth relative to the rings, $\omega_r$, is positive when the rings are tilted from ecliptic towards the observer and negative when the rings are tilted from the ecliptic away from the observer and changes from -90\deg\ to 90\deg.
We do not consider other possible values of 
$\omega_r$=$\pm$($90-180$)\deg\ because these orientations produce light curves symmetric to those with $\omega_r$=$\pm$($0-90$)\deg.
The geometry is fully determined by $i$ and $\Theta(t)$ for a non-ringed planet and by $i$, $\Theta(t)$, $\omega_r$, and $\epsilon$ for a ringed planet.
}
\label{fig:angles}
\end{figure} 
The only time-dependent parameters are the orbital angle $\Theta$ and phase angle $\alpha$.
For a non-ringed planet the star-planet-observer geometry is fully determined by two angles $i$ and $\Theta$ ($\alpha$ is linked to $i$ and $\Theta$).
For a ringed planet the geometry is determined by four angles: $i$, $\Theta$, $\omega_r$ and $\epsilon$.} 

\subsection{Reflecting properties of Jupiter, Saturn and rings}
\begin{figure}[tbp]
\plotone{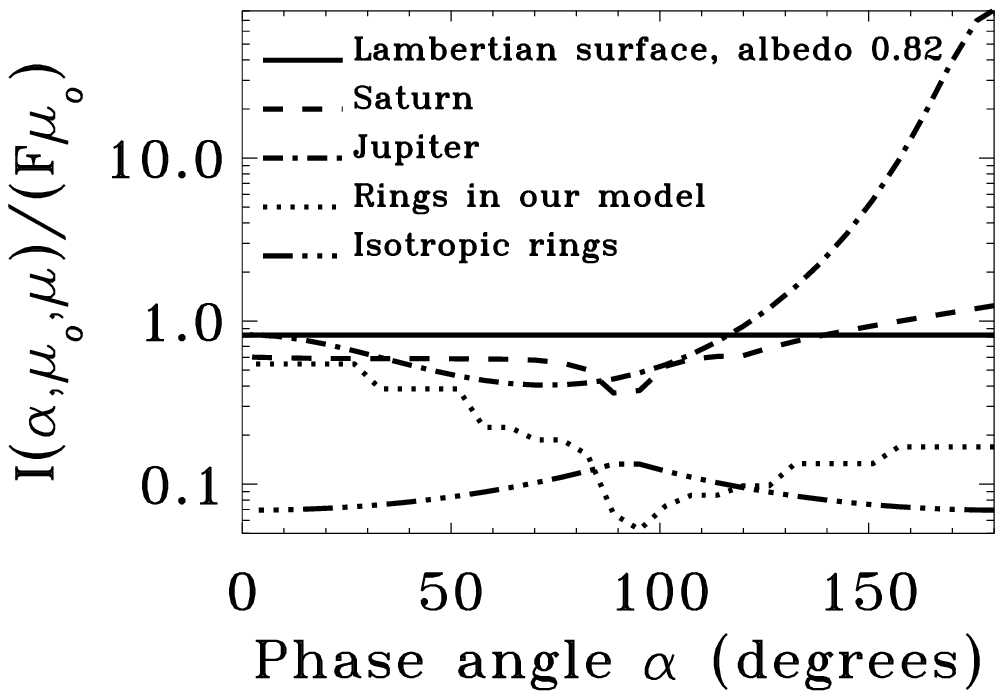}
\vspace{-2in}
\caption{Scattering phase functions for a Lambertian surface, Saturn, Jupiter, Saturn's rings (as used in our model), and the phase function for rings consisting of isotropically scattering particles used by \cite{arnold04}.
The phase functions are plotted for a single sample geometry: $\mu_0$ is fixed (the Sun is  2\deg$ $ above horizon), while the observer moves in the plane that includes the Sun and zenith.
The wavelengths correspond to visible red light (0.6-0.7\mm). 
}
\label{fig:phase_functions}
\end{figure}

\corc{Figure \ref{fig:phase_functions} shows the scattered brightness, for a given geometry, derived from our model for the surfaces of Jupiter, Saturn, and its the rings in red wavelengths (0.6-0.7\mm)}. 
Note the logarithmic scale of the ordinate and the large amplitudes of the phase functions.
The normalized brightness $I(\alpha,\mu_0,\mu)/(F\cdot\mu_0)$ of a point on the surface 
is plotted versus scattering phase angle $\alpha$.
$F\cdot\mu_0$ is the ``ideal'' reflected brightness of a white isotropically scattering (Lambertian) surface, where $\mu_0$ is the cosine of the incidence angle 
measured from the local vertical.  
$F\cdot(\pi \rm{~ steradians})$ is a solar flux at the planet's orbital distance. 
In our notation, $\alpha=0$\deg\ indicates backward scattering 
and $\alpha=180$\deg\ indicates forward scattering.
Lambertian scattering with reflectivity 0.82 (matching Saturn's full-disk albedo at opposition) is shown as a horizontal solid line for comparison.
We also indicate, for the same ring opacity, albedo and geometry,  the ring phase function used by \cite{arnold04}, 
which assumes isotropically-scattering particles.

For Jupiter, we assume that the scattering phase function depends only on $\alpha$.
\corc{For Saturn with rings, the scattering phase function depends on three angles: incidence (via $\mu_0$), emission (via $\mu$), and phase (via $\alpha$), and so cannot be described fully by the one-parameter function of $\alpha$.}
Figure \ref{fig:phase_functions} shows reflected light only for the geometry in which the Sun is 2\deg$ $  above the horizon ($\mu_0$=0.035) and the observer moves from the Sun's location ($\alpha=0$\deg) across the zenith toward the point on the horizon opposite to the Sun ($\alpha=178$\deg). 
The small amplitude for isotropically scattering rings in Fig. \ref{fig:phase_functions} is due to the very grazing illumination chosen for this example.  
At less grazing illuminations, the amplitudes for isotropic rings and our rings are more similar.

\subsubsection{Jupiter}

For each point on Jupiter, the strong forward scattering of the reflecting clouds is represented by the two-term Henyey-Greenstein function $P(g_1,g_2,f,\alpha)$.
\begin{equation}
I(\alpha,\mu_0,\mu)/F={\mu_0}\cdot P(g_1,g_2,f,\alpha)
\end{equation}
\begin{equation}
P(g_1,g_2,f,\alpha)=fP_{HG}(g_1,\alpha)+(1-f)P_{HG}(g_2,\alpha)
\label{eq:h_g}
\end{equation}
The individual terms are Henyey-Greenstein functions representing forward and backward scattering lobes, respectively.
 \begin{equation}
P_{HG}(g,\alpha)=2\cdot{{(1-g^2)}\over{(1+g^2+2g \cdot \cos{\alpha})^{3/2}}}~~,
\end{equation}
where $\alpha$ is the phase angle , $f$ is the fraction of the forward versus backward scattering, and $g$ is one 
of $g_1$ or $g_2$;  
$g_1$ is positive and controls the sharpness of the forward scattering lobe and $g_2$ is negative and controls the sharpness of the backscattering lobe.
Note that we have rewritten $P_{HG}$ relative to common notation 
in which the scattering angle $\theta=180$\deg$-\alpha$ is used; this reverses the sign in front of the $2g \cdot \cos{\alpha}$ term.
Unlike the common use of the Henyey-Greenstein function for single-particle scattering, in which the function is normalized over the sphere of solid angles around the particle, we use this function only as a convenient analytical expression to fit the measured scattering of Jupiter's surface.
In this case, the function is not normalized over the hemisphere of solid angles above the surface.
The 
\corb{spherical} 
albedo (the ratio of reflected to incident light for the whole planet 
\corb{at appropriate wavelengths}) 
is imbedded in the scattering function, and is a rather complicated integral over the planet's surface that we do not calculate here. 

\begin{figure}[tbp]
\plotone{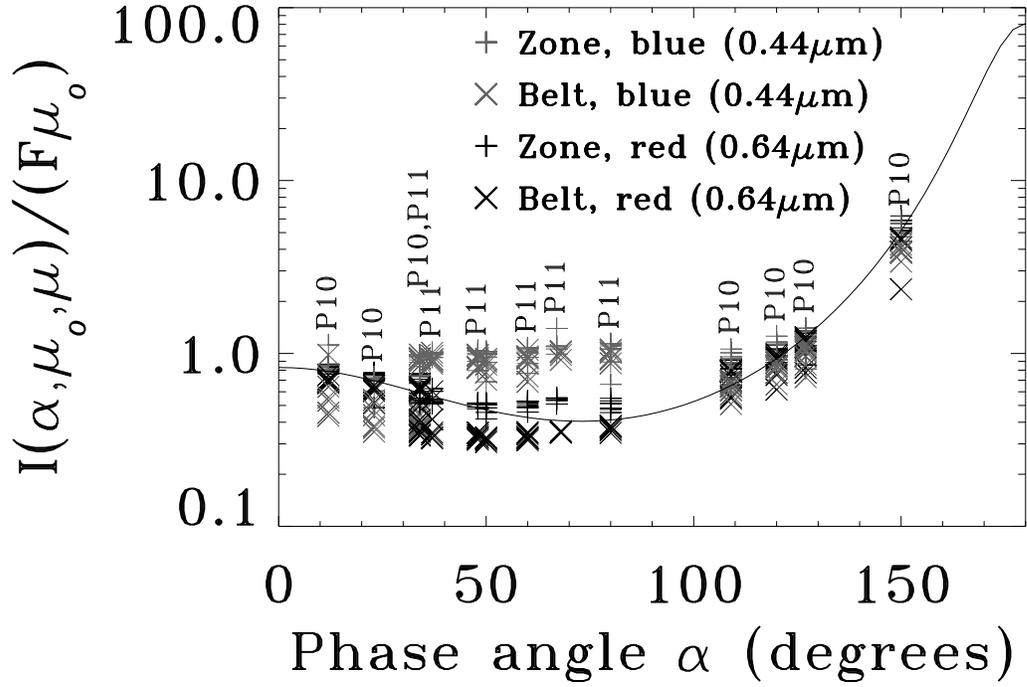}
\vspace{-2in}
\caption{
\corc{Our fit of the Henyey-Greenstein function (solid line, $g_1=0.8$, $g_2= - 0.38$, $f=0.9$) to the Pioneer 10 data (labeled P10) published as Tables IIa and IIb of \cite{tomasko78}, and Pioneer 11 data (labeled P11) published as Tables IIa and IIb of  \cite{smith84}. 
The data represent belts (dark stripes) and zones (bright stripes) on Jupiter observed with the red (0.595--0.720 \mm) and blue (0.390--0.500 \mm) filters.}
}
\label{fig:tomasko78fit}
\end{figure}

Figure \ref{fig:tomasko78fit} shows our fit of the 
Henyey-Greenstain function, with $g_1=0.8$, $g_2 = -0.38$, and $f=0.9$, 
to the data points from the Pioneers 10 and 11 images \citep{tomasko78,smith84}. 
\corc{The values we have chosen for $g_1$, $g_2$, and $f$ also reproduce the full-disk albedos in the spectra observed by \cite{karkoschka94} and \cite{karkoschka98} at the wavelengths corresponding to the red passband of Pioneer.}
Pioneer images are taken with broadband blue (0.390--0.500 \mm) and red (0.595--0.720 \mm) filters at different locations on Jupiter.
In particular, \cite{tomasko78} and \cite{smith84} have indicated two types of locations: the belts, usually seen as dark stripes on Jupiter (\tms\ symbols on the plot), and zones, usually seen as bright stripes on Jupiter (+ symbols on the plot).
The relative calibration between Pioneer 10 and Pioneer 11 data is not as well constrained as the calibration within each data set.

Accepting the calibrations given in \cite{tomasko78} for Pioneer 10  and \cite{smith84} for Pioneer 11, our model curve better represents the observations in the red filter (black data points) than in the blue.
We assume here that this curve represents the red wavelengths.
The Pioneer 11 blue data at moderate $\alpha$ in Fig. \ref{fig:tomasko78fit} seem to be systematically offset from the Pioneer 10 points, 
which may be a result of relative calibration error. 
\corc{Consequently, we do not model the blue wavelengths.}

The vertical spread of the points in the same filter and at the same location is due mainly to differences in emission and incidence angle between the points.
This spread is not large, \ie, the reflectivity is not strongly dependent on incidence and emission angles for fixed $\alpha$ other than a $1/\mu_0$ dependence, which is often labeled ``Lambertian limb-darkening.''
\corc{We note, however, that the Pioneer-measured phase angle dependence is strongly non-Lambertian, which has important consequences for the light curves we generate.} 

In addition to Pioneer data, images of Jupiter from a variety of angles were taken by Voyager, Galileo, and Cassini, though we are not aware of any other published data of the scattering phase functions for the Jovian 
surface.\footnote{One of us, UD, plans to work on obtaining the spectral phase functions from Cassini nine-filter visible images in the immediate future.}
Data for $\alpha>$150\deg$ $ do not exist because these directions would risk pointing spacecraft cameras too close to the Sun; this limits the ability of the data to constrain the forward-scattering peak at $\alpha\sim$180\deg. 
Our derived light curves are not severely affected by this uncertainty, however, since when the forward scattering is important, the observed crescent is small, and reflected light phase curve undergoes its minimum. 

\subsubsection{Saturn}
The scattering phase function and albedo of Saturn are represented by the Backstorm law, which was used by \cite{dones93} to fit Saturn's data.
\begin{equation}
I/F={A \over \mu} \left({{\mu\cdot \mu_0}\over{\mu+\mu_0}}\right)^B~~,
\end{equation}
where $\mu$ and $\mu_0$ are the cosines of emission and incident angles measured relative to the local vertical, respectively, 
and $A$ and $B$ are coefficients that depend on the phase angle $\alpha$.
The coefficients are fitted by \cite{dones93} to Pioneer 11 fitted phase function tables, which were produced by the multiple scattering model of \cite{tomasko84}.
We averaged the coefficients published by \cite{dones93} separately for zones and belts through Pioneer's blue and red filters. 

Table \ref{tab:backstorm} gives the resulting coefficients; Fig. \ref{fig:phase_functions} displays the red filter curve only. 
\begin{table}[htbp]
  \begin{center}
       \begin{tabular}{|c|c|c|c|c|c|c|c|c|}
\hline
Phase angle $\alpha$            &0\deg&30\deg&60\deg&90\deg &120\deg &150\deg &180\deg\tablenotemark{a}\\
\hline
A (red, 0.64 \mm)       &1.69   &1.59    &1.45  &1.34   &1.37   & 2.23  &3.09\tablenotemark{a}\\
B (red, 0.64 \mm)       &1.48   &1.48    &1.46  &1.42   &1.36   & 1.34  &1.31\tablenotemark{a}\\
A (blue, 0.44 \mm)      &0.63   &0.59    &0.56  &0.56   &1.69   & 1.86  &3.03\tablenotemark{a}\\
B (blue, 0.44 \mm)      &1.11   &1.11    &1.15  &1.18   &1.20   & 1.41  &1.63\tablenotemark{a}\\
\hline     
      \end{tabular}
    \caption{Coefficients for the Backstorm function for Saturn, averaged between belt and zone values published in Table V of \cite{dones93}.}
\tablenotetext{a}{The coefficients at $\alpha$=180\deg$ $ are not constrained by observations; and were estimated by linearly extrapolating the coefficients at 120\deg$ $ and 150\deg} 
    \label{tab:backstorm} 
  \end{center}
\end{table}

The Pioneer spacecraft did not observe Saturn at $\alpha>$ 150\deg. 
Corresponding coefficients in \cite{dones93} are given for $\alpha$ up to 150\deg.
To complete the phase curve, we extrapolate the coefficients at 180\deg$ $ linearly from their values at 120\deg$ $ and 150\deg.
The phase curve given in Fig. \ref{fig:phase_functions} 
corresponds to the case when the incidence angle is fixed at 88\deg\ (the Sun is 2\deg\ above the horizon,\ie, $\mu_0$=0.035) and the trajectory of the observer on the sky is in the plane including zenith and Sun's location ($\mu$ is linearly dependent on $\alpha$). 
This cross-section of the fitted three-parameter phase function is not as smooth as the Henyey-Greenstein function for Jupiter.
For example, the dip at $\alpha=90$\deg$ $ is a result of the approximation by an analytic function rather than real observations.

\begin{figure}[htbp]
\plotone{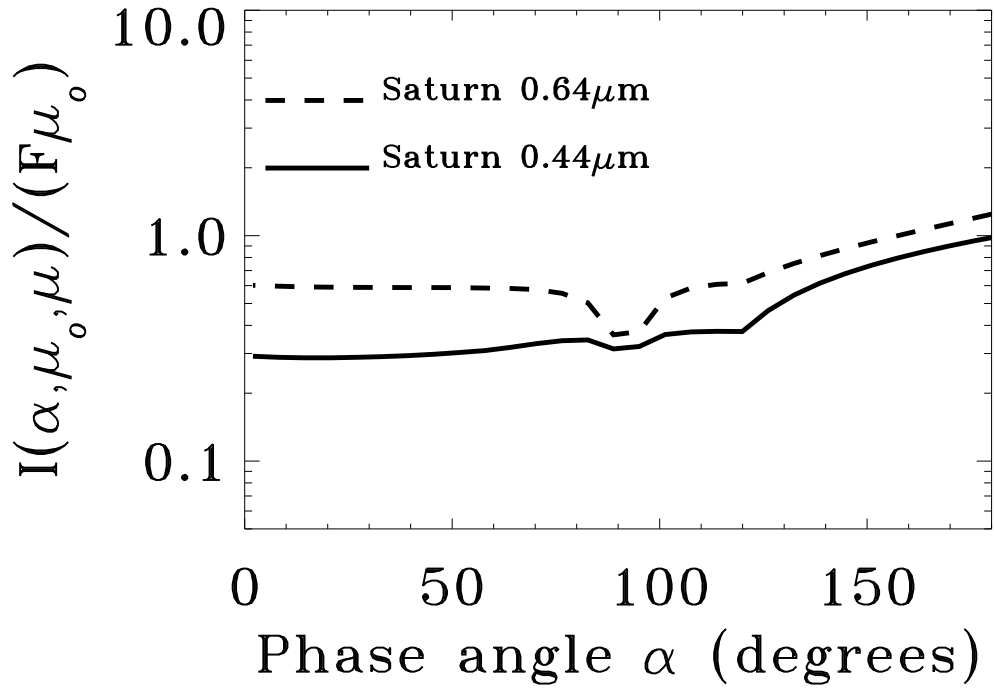}
\vspace{-2in}
\caption{Scattering phase functions for Saturn in red ($0.595-0.720$ \mm) and blue ($0.390-0.500$ \mm) passbands adopted from \cite{dones93}.}
\label{fig:phase_functions_saturn_red_blue}
\end{figure}

Figure \ref{fig:phase_functions_saturn_red_blue} demonstrates the difference between the blue and red phase functions for Saturn.  
The values for the blue curve are $1.5-2$ times smaller, indicating that Saturn is darker in blue wavelengths 
\corb{due to higher absorption by photochemical hazes.}

\corb{The ring brightness in reflection (top-side) and transmission (bottom-side) is provided by a physical scattering model of the ring.
This model calculates multiple scattering within the rings using a ray tracing code.  
The model ring is populated with macroscopic bodies of 1m-size, with optical depth and albedo profiles chosen to match those of \cite{dones93}.
Note that \cite{dones93} found the amount of dust in the main rings to be \corc{too} small to contribute substantially to ring reflection.  
The model reproduces well the brightnesses observed by Voyager 1 and 2.  
We use the code to predict ring brightness at geometries not observed by Voyager. 
The  opacity for each pixel is taken at the point on the rings that projects to the pixel's center.}

\corb{The ring brightness at each point is a function of three angles ($\alpha,\mu,\mu_0$), the  optical depth, and the albedo.  
Due to data volume restrictions we have binned output in rather large steps, which depend on parameter values.
These steps are clearly seen in the ring phase function (dotted line) in Fig. \ref{fig:phase_functions}.
The phase function in Fig. \ref{fig:phase_functions} corresponds to the same scattering geometry as the curves for Saturn shown in Fig. \ref{fig:phase_functions_saturn_red_blue}.
The optical depth at the sample point on the rings (1.7 times Saturn's radii from the planet's center) is 2.3, the albedo of the ring particles is 0.56.
The ring is observed from the illuminated side.
}

To produce light curves of the fiducial exoplanets that we model, images of the planet for a set of locations along the orbit are generated.

For each image we integrate the total light coming from the planet and the rings (if any) to obtain the full-disk 
\corb{(or geometric)} 
albedo $p(\alpha)$.
\begin{equation}
p(\alpha)={\sum_{\rm pixels} I(\mu,\mu_0,\alpha)
\cdot r_{\rm pix}^2/F \over \pi R_P^2}~~~,
\label{eq:pixel_integral}
\end{equation}
where $r_{\rm pix}$ is the pixel size and $R_P$ is the planet's radius.
The full-disk albedo is the planet's luminosity $L_P$ normalized by the reflected luminosity of a Lambertian disk with the planet's radius at the planet's orbital distance, illuminated and observed from the normal direction.
\begin{equation}
p\equiv L_P/(\pi R_P^2 F)
\label{eq:full_disk_albedo_definition}
\end{equation}
\corb{Generally, $I$, $L_P$, $F$, and $p$ depend on wavelength.
However since we have integrated over the red passband, these values are weighted averages over the Pioneer filter at red wavelengths.}


\subsection{Eccentric Orbits}
\label{sec:model_of_eccentric_orbit}
In addition to modeling light curves from \cora{ringed} planets traveling in circular orbits, we also model the light curves of \cora{non-ringed} planets moving in \cora{eccentric} orbits.  Although most of the planets in \corc{the Solar system} 
orbit the Sun with low
eccentricities, the extrasolar planets detected to date display a wide range of eccentricities\footnote{Eccentricities of extrasolar planets are listed at http://exoplanets.org/almanacframe.html}.

Eccentric orbits differ from circular orbits in two main respects.  First, the distance between the star and the planet varies according to the
planet's orbital position.  As the planet moves from its pericentre to its
apocentre, its separation from its host star increases.  
Second, in an eccentric orbit, 
the planet does not move with a constant angular speed around the
star as it would in a circular orbit.  The planet's angular speed varies
throughout the orbit, being a maximum at the pericentre and minimum at the apocentre.  

To model the reflected light as a function of time, 
we calculate the angular
position of the planet and the planet-star separation over a
complete period using a solution to Kepler's equation (in notation of \cite{murray00b}):      
\begin{equation}
\label{eq:kepler}
M = E-e\,{\sin}E 
\end{equation}
where M is the mean anomaly (a parameterization of time), $E$ is the
eccentric anomaly (a parameterization of the polar angle) and $e$ is the
eccentricity.  
Kepler's equation is transcendental in $E$ and therefore cannot be
solved directly.  
\corc{Since we need only one complete orbit to model a light curve, 
however, we are able to use the simple Newton-Ralpson method to obtain a numerical solution to Kepler's equation.}  
Applying the Newton-Ralpson method\footnote{Application of Newton-Ralpson method to Kepler's equation is described in \cite{murray00b}} to
Eq. \ref{eq:kepler} we \corc{obtain the iterative solution}:
\begin{equation}\label{eq:newton}
E_{i+1} = E_{i}-\frac{E_i-e\, {\sin}E_i-M_{i}}{1-e\, {\cos} E_{i}} 
\end{equation}
Using this iteration, we calculate the angular position and planet-star
separation as functions of time, and use this to model the amplitude of reflected light from the planet.

\begin{figure}[tbp]
\vspace{-4in}
\plotone{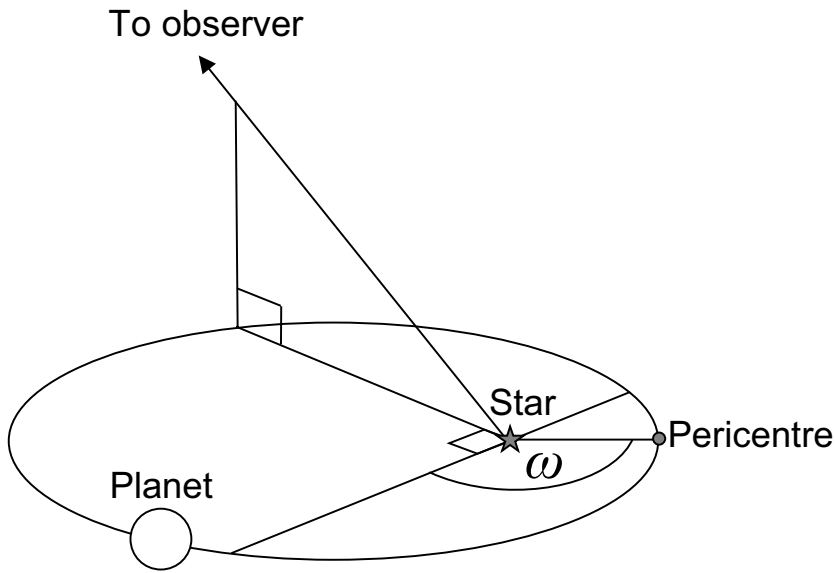}
\vspace{-1.5in}
\caption{
\corc{The argument of pericentre, $\omega$, is an additional angle needed to define geometry of a non-ringed planet on eccentric orbit.} 
}
\label{fig:angles_eccentric}
\end{figure}

As can be seen in Fig. \ref{fig:angles_eccentric}, the geometry of an ellipse \corc{introduces an additional parameter, 
in addition to inclination}, in the observer's azimuthal perspective of the system.  
This parameter is the argument of pericentre $\omega$, which is the angle (in the planet's orbital plane) between the ascending node line and pericentre \citep{murray00b}. 
\corb{The reference plane is formed by the observer's line of site and its normal in the ecliptic plane, which is also the accending node line.}
\corb{Viewing the system from the line of the pericentre 
(planet's smallest phase is at pericentre) would give argument of pericentre 90\deg.
Viewing the system from the line of the apocentre (planet's largest phase is at pericentre) would give argument of pericentre $-90$\deg.}

\section{Results}
\label{sec:results}

We modeled several geometries for the planetary system, \ie, different orbital eccentricities $e$ for non-ringed planets, and different ring obliquities $\epsilon$ relative to the ecliptic for ringed planets on circular orbits.
For each of these cases we modeled a variety of observer 
locations, \ie, different orbital inclinations $i$ as 
seen by the observer, and different azimuths $\omega$ of the observer relative to the orbit's pericentre or to the rings ($\omega_r$). 
In what follows, we compare light curves for these differing 
geometries, and discuss whether or not different
geometric effects can be distinguished from one another 
from the characteristics of the light curve alone.

\subsection{Jupiter versus a ringless Saturn}
\label{sec:jupiter_saturn}

We first tested how light curves would differ for (ringless) 
exoplanets with the surface reflection characteristics 
of Jupiter and Saturn.
We modeled the planets for a variety of inclinations, 
assuming circular orbits.
Low-inclination orbits (close to face-on) produce smaller light curve amplitudes because the planet appears 
to be at half-phase most of the time.
The most prominent amplitudes, and the most prominent difference between Jupiter and Saturn reflectance properties, 
occurs for edge-on orbits, for which the phase changes from completely ``new moon'' to completely ``full moon''.

\begin{figure}[htbp]
\vspace{-1in}
\plotone{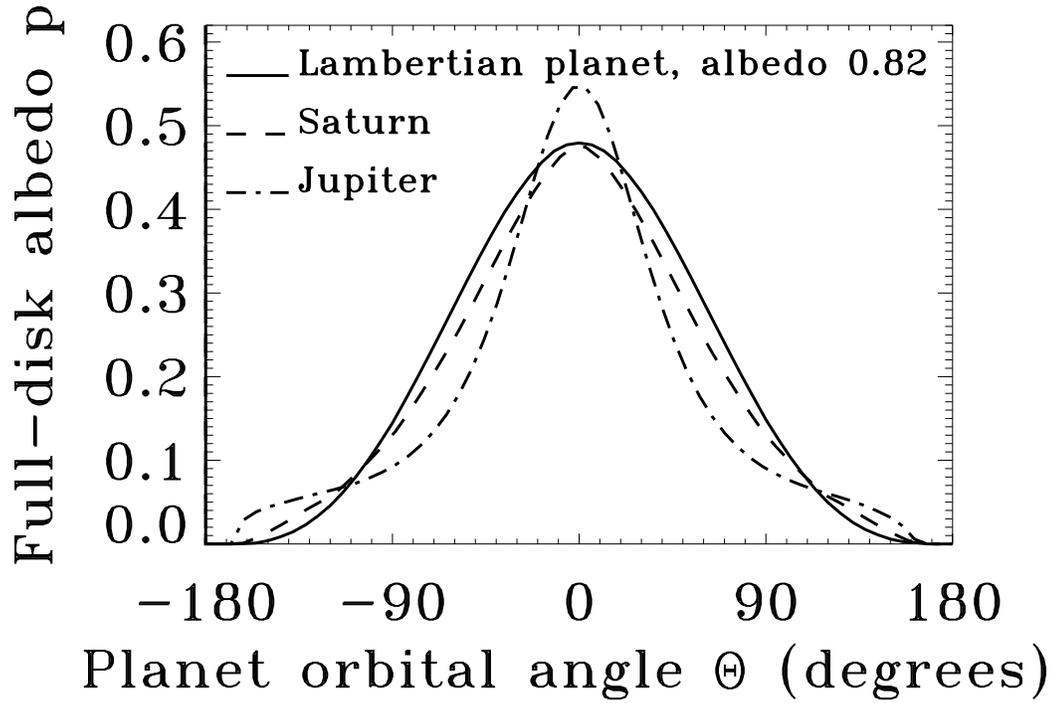}
\vspace{-2in}
\caption{Comparison of light curves for a spherical Jupiter, a spherical Saturn , and a spherical Lambertian (isotropically scattering) planet, assuming the planets are ringless 
and have the same radii $R_P$. The planets differ only in their  surface scattering properties.
Albedos are shown for visible red light (0.6-0.7\mm).  
}
\label{fig:jup_sat}
\end{figure}

Figure \ref{fig:jup_sat} compares edge-on light curves for a 
ringless Saturn, Jupiter, and a Lambertian planet.
The luminosity of the planet is normalized by the incident stellar illumination to obtain the full-disk albedo $p$ as described in equations (\ref{eq:pixel_integral}) and (\ref{eq:full_disk_albedo_definition}).
The planets are spherical and have identical radii; 
the curves differ only because of the 
different surface scattering of the three planets.

The full-disk albedo can be converted into the planet's luminosity $L_P$ as a fraction of star's luminosity $L_*$ for a planet of equatorial radius $R_P$ at an orbital distance $D_P$.
\begin{equation}\label{eq:lp_l*}
L_P/L_*=(R_P/D_P)^2\cdot p
\end{equation}
For example, for Saturn at 1 AU, 
$(R_P/D_P)^2\approx$1.6\tms$10^{-7}$.
The abscissa of Fig. \ref{fig:jup_sat}
indicates the azimuthal angle of the planet in its orbital plane, or the orbital angle $\Theta$ (see Fig. \ref{fig:angles}), starting at minimum planet phase ($\Theta=-180$\deg).
Note that only for edge-on inclinations 
does the minimum phase along the orbit correspond 
to a completely ``new moon''.
For orbits with lower inclinations, the minimum phase would be a crescent of a size between ``new moon'' and ``half moon''.
The corresponding maximum phase will be between the ``full moon'' and ``half moon'' at $\Theta=0$\deg.
For circular orbits, the plot can be transformed into a 
time-dependent light curve simply by dividing 
$\Theta$ by 360\deg\ and multiplying by the planet's orbital period.

Study of the spectral dependence of reflected light curves 
is beyond the scope of this paper; \corc{we will 
address this is a future paper}. 
We do note, however, that Fig. \ref{fig:tomasko78fit} 
indicates that at shorter wavelengths Jupiter 
is less forward scattering than at longer wavelengths.
Also, Figs. \ref{fig:tomasko78fit} and \ref{fig:phase_functions_saturn_red_blue} 
indicate that both Jupiter and Saturn are darker in blue 
than in red at observational geometries close to
back scattering ($\alpha\approx$ 0\deg).

The light curve for Jupiter peaks much more sharply 
at full phase ($\Theta\approx$ 0\deg) than the light curves 
of Saturn or the Lambertian planet 
because of the sharp back-scattering peak in 
Jupiter's scattering phase function 
($\alpha\approx$ 0\deg$ $ in Fig. \ref{fig:phase_functions}).
Near zero phase ($\Theta\approx\pm$180\deg$ $ in Fig. \ref{fig:jup_sat}), Jupiter is more luminous 
than the other two models 
due to the large forward scattering from its surface ($\alpha\approx$ 180\deg$ $ in Fig. \ref{fig:phase_functions}).

Such differences in phase functions is commonly attributed to larger particle size for the main cloud deck on Jupiter.  
For example, \cite{tomasko78} suggest particle sizes larger than  0.6 \mm to explain the forward scattering.

\subsection{Oblate Planets}
\label{sec:oblate_planet}
\corb{Saturn's equatorial radius is 10\% larger than its polar radius.
This 10\% oblatenes of Saturn (6\% for Jupiter) makes the planet appear larger when looking from the pole than when looking from the equator, which affects the light curves.}

\corb{Figure \ref{fig:oblateness} compares light curves for a spherical planet, a 10\% oblate planet viewed from its equator, and a 10\% oblate planet viewed from 45\deg\ latitude.
\begin{figure}[htbp]
\vspace{-1in}
\plotone{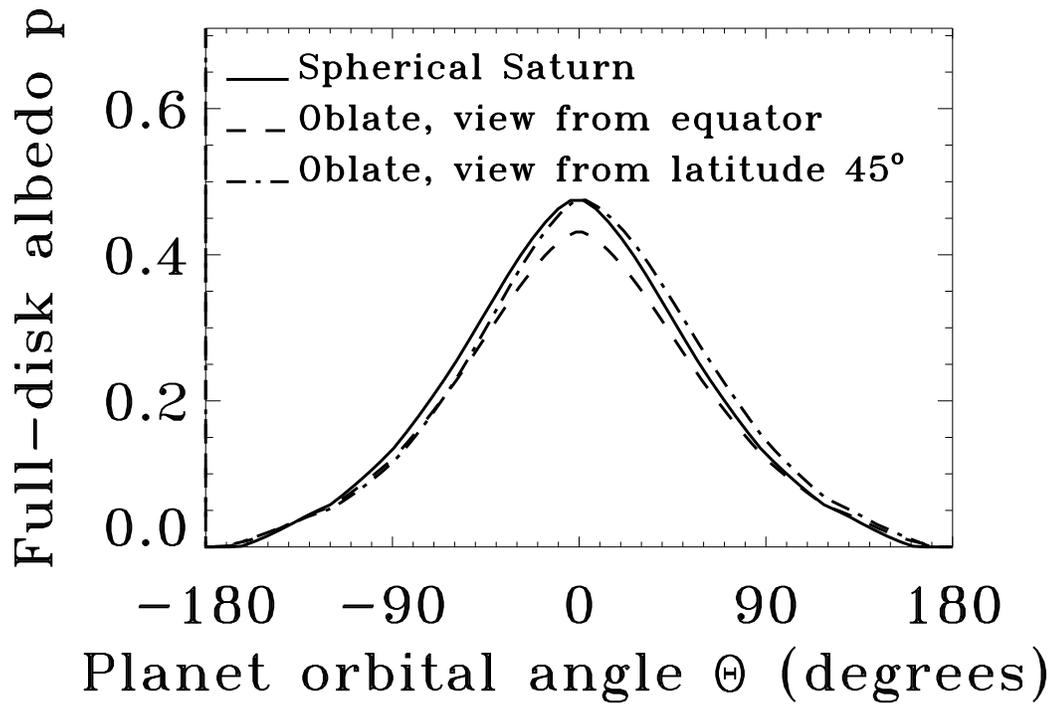}
\vspace{-2in}
\caption{Comparison of light curves for a spherical and a 10\% oblate planet with Saturn's surface properties.
Planets have the same equatorial radii and are ringless.
The orbit is observed edge-on ($i$=90\deg).
The planets are rotating ``on their side'' ($\epsilon$=90\deg). 
Albedos shown are for visible red light (0.6-0.7\mm).
}
\label{fig:oblateness}
\end{figure}
\corc{All three sample planets have the same equatorial radius and scattering properties of Saturn.  The light curves in 
are Fig. \ref{fig:oblateness} are 
displayed for an edge-on orbit ($i$=90\deg) of a planet rotating ``on its side'' ($\epsilon$=90\deg), because this 
geometry emphasize the effects of oblateness in the 
reflected light. 
The differences among the curves is created by differences in the observer's azimuth relative to the planet's equator. (This angle is analogous to $\omega_r$ but is measured with 
respect to the equatorial plane rather than the ring plane.
The rings are in the equatorial plane for Saturn and thus the same $\omega_r$ characterizes the observer's azimuth.)
Observing the planet from the equator decreases the cross-section of the planet and thus decreases the amplitude of the curve.
Observing the planet from the pole yields a curve indistinguishable from the solid curve for a spherical planet;
this is not shown in Fig. \ref{fig:oblateness}.
Observing the planet from latitude 45\deg\ produces a small asymmetry in the curve because on one side of the orbit the flatter (and larger in projection) pole is better illuminated than on the other side of the orbit. 
This asymmetry is small compared to the asymmetries produced by rings, as we discuss in the next section. }}

\subsection{Ring effects}
\label{sec:ring_effects}

The presence of rings has a large effect on the light curves 
of planets in reflected light.
In order to describe the geometry of ringed planetary systems, 
two more parameters are required; even assuming a fixed radial density distribution for the rings, the geometry 
becomes sensitive to the ring obliquity $\epsilon$ (the angle between the ring and ecliptic planes), and to the azimuth of the observer relative to the rings $\omega_r$ (see Fig. \ref{fig:angles_eccentric}).
Together with the variable orbital inclination $i$, this produces 
the three-parameter space that we have sampled to explore possible light curves for ringed exoplanets similar to Saturn.

\begin{figure}[htbp]
\plotone{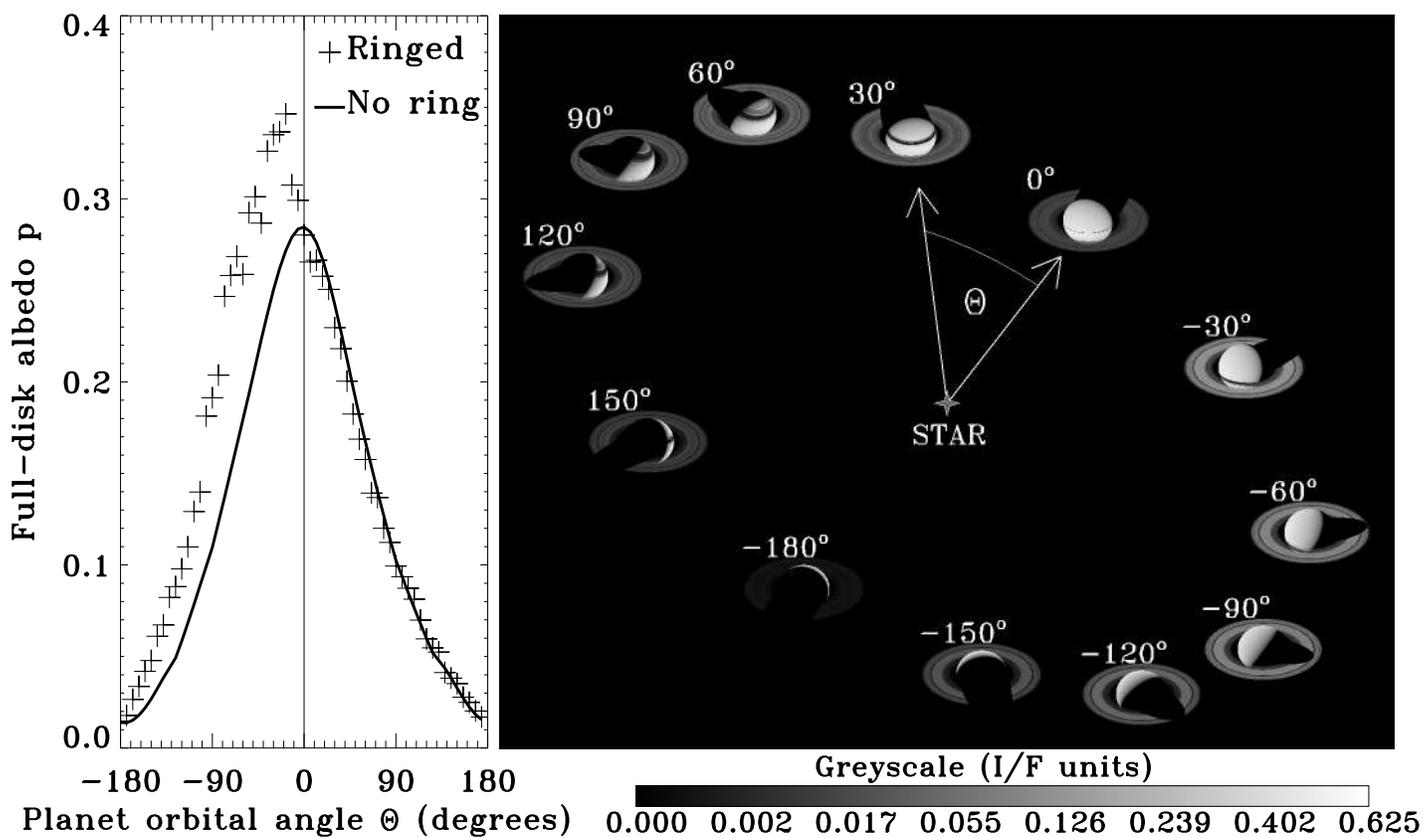}
\vspace{-1.3in}
\caption{Effect of Saturn's rings if observed at 35\deg$ $ above the orbital plane ($i$=55\deg). 
In this example, Saturn's ring's obliquity is $\epsilon$=27\deg.
The observer's azimuth on the ecliptic is separated from the intersection of the ring and ecliptic planes by $\omega_r$=25\deg.
The plot on the left shows light curves for a ringless planet (solid curve) and for a ringed planet (crosses).
Values of $\Theta$ are indicated next to each image. 
The brightness of the images in the cartoon is given in the units of $I/F$, as indicated on the nonlinear greyscale bar.
}
\label{fig:cartoon}
\end{figure}

Figure \ref{fig:cartoon} shows an example light curve for $i$=55\deg, $\epsilon$=27\deg\ and $\omega_r$=25\deg, which is a convenient geometry to demonstrate different ring effects in one light curve.
The cartoon on the right of Fig. \ref{fig:cartoon}
displays images of Saturn at several positions on the orbit as it 
would appear to a remote observer.
The plot on the left shows the light curve for the ringed planet (+ symbols) compared to the curve for the same geometry 
for a ringless planet (solid curve). 
The $\sim$10\% spread of the points in the ringed light curve 
is due partially to the large steps in the ring reflection table (see Fig. \ref{fig:phase_functions}) and should be treated as a model uncertainty.
For a large fraction of the orbit, the ring presence increases the luminosity of the planet ($\Theta<-20$\deg) because 
the observer is able to view more reflecting surface.
However for $0 < \Theta < 100$\deg$ $
the rings are not well illuminated and in fact shadow part of the planet, 
producing a lower luminosity than for the non-ringed planet.
More importantly, the ringed light curve is asymmetric, which makes it distinguishable from any light curves of a non-ringed planet on a circular orbit.
Obviously, planets on eccentric orbits may produce asymmetric curves due to variation in planet-star separation along the orbit.
Light curves for non-ringed planets on eccentric orbits thus could 
be confused with those of ringed planets in the absence of radial velocity data (see Section \ref{sec:results_eccentric_orbit} for discussion).

\begin{figure}[htbp]
\plotone{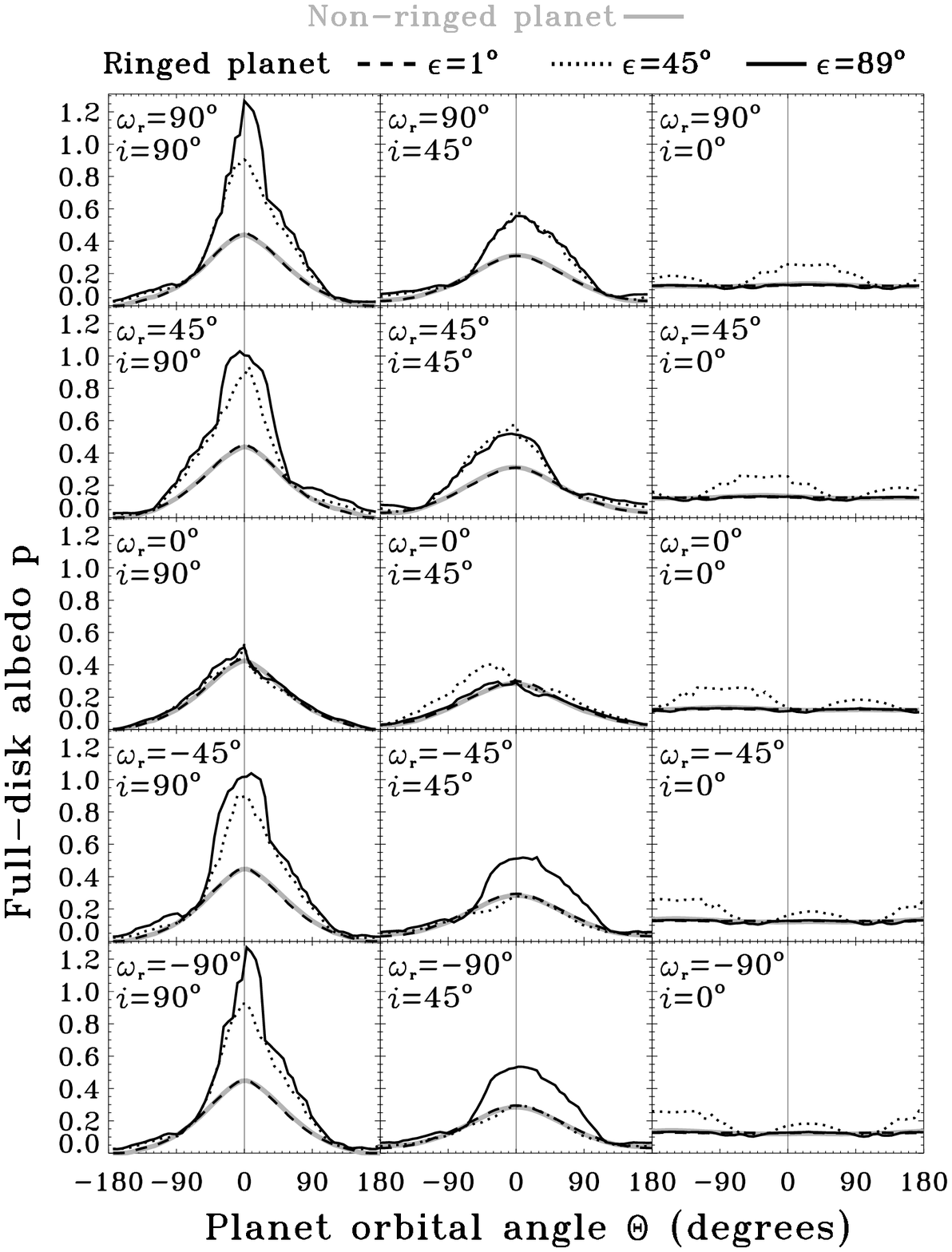}
\vspace{-0.8in}
\caption{
Light curves for Saturn for different geometries.
Different ring obliquities $\epsilon$ are shown on each 
subplot as black curves. 
A curve for a spherical planet without rings is shown on each subplot in grey. 
Each column corresponds to a different orbital inclinations $i$.
Each row corresponds to a different azimuth 
$\omega_r$ of the observer relative to the rings. 
}
\label{fig:ring_summary}
\end{figure}

The variety of light curves for a Saturn-like planet at different geometries is illustrated in Figure \ref{fig:ring_summary}.
In this figure, different ring's obliquities are shown in each 
subplot as black curves.
The solid black curve indicates an obliquity of $\epsilon$=89\deg$ $ (that is, the planet is rotating on its side, and rings are located in the equatorial plane and are nearly perpendicular to ecliptic).
The dotted black curve is for $\epsilon$=45\deg, and the dashed 
black curve is for $\epsilon$=1\deg$ $ (rings are nearly in the ecliptic plane).
Grey curves illustrate the same geometries for a non-ringed planet.

Each subplot within Fig. \ref{fig:ring_summary}
corresponds to a different observer geometry.
In the left column, the orbit is observed edge-on ($i$=90\deg); 
the middle column corresponds to an observer 45\deg$ $ above the ecliptic ($i$=45\deg); the 
right column shows the orbit observed face-on ($i$=0\deg).
Each row corresponds to different azimuths of the observer relative to the rings $\omega_r$. The top row illustrates geometries with 
$\omega_r$=90\deg.
\corc{At this azimuth, the rings are seen at the maximum possible opening, assuming all other geometric parameters (ring obliquity $\epsilon$ and orbital inclination $i$) are fixed.
In lower rows, the observer's azimuthal is rotated with respect to the orbital plane, resulting in a more grazing view of the rings.
The bottom row is for $\omega_r$=90\deg, for which the rings are seen at the most grazing geometry possible for fixed  $\epsilon$ and $i$.}

Several lessons can be learned from Fig. \ref{fig:ring_summary}. 
First, rings generally 
increase the amplitude of the light curves by a factor of two to three.
Recall that precise, but unresolved observations will only 
be able to detect the variations on top of the large 
constant flux of the star.  
An amplitude increase due to rings could be partially 
confused with effects of due to a larger planet size or albedo.
The ambiguity may be resolved spectrally because the spectrum 
of rings should be rather flat, whereas the planetary atmosphere is expected to be dark at a set of prominent gaseous absorption bands.

The second lesson 
is that light curves for a ringed planet display two types of 
asymmetry.
The first, and potentially easier to detect, asymmetry 
is the offset of the curve's maximum relative to $\Theta=0$\deg, the maximum point for light curves 
of a ringless planet (marked by vertical lines in Fig. \ref{fig:ring_summary}).
This offset of the light curve maximum itself occurs in rather small fraction of the plots.  \corc{However, since the entire curve is 
asymmetric, the effect would be detectable as an overall deviation 
from the simple symmetric curves likely to be used to fit 
the first detections of reflected light. 
If radial velocity data exist for such a planet, the exact timing of the $\Theta=0$\deg$ $ point and the eccentricity of the orbit would be 
measurable.
In this case, a shifted maximum in the light curve would 
yield direct evidence of the rings when compared to the asymmetry 
expected for any orbit eccentricity.}

The offset of the light curve maximum due to rings is a strong photometric signature that was also noted in the 
ring simulation of \cite{arnold04} (as a $\phi-$shift).
We stress here, however, that other processes may be 
capable of producing such a $\phi-$shift.
Similar, although probably smaller, shifts may be induced by seasonal brightness variation on the planet.
None of the giant planets in the Solar System have pronounced seasonal brightness variations, but extrasolar planets may display seasons, in which case the variations induced by them may cause 
an offset of the brightness maximum.
Offsets could also be produced by a global asymmetry 
of the brightness distribution over the planet.
An example in the Solar System is the striped appearance of Jupiter.
Jupiter's  brightness varies with latitude by as much as 50\% in  broadband visible light \citep{smith84}.
A reasonable scenario in which an extrasolar planet's poles are brighter than its equator would cause the planet to appear globally brighter in winter and summer, and darker at mid-seasons, which could induce 
a $\phi-$shift.
\corb{Oblateness of a planet may also produce a small  $\phi-$shift (see Section \ref{sec:oblate_planet}).}
Again, spectroscopy may help to resolve the ambiguity between 
asymmetries caused by rings and by processes on the planet surface.

Asymmetry is also apparent in the shape of the light curve, 
but resolving such detail would require another order of magnitude in the instrument sensitivity.
If this fine structure of the light curves could be measured, 
however, detection of rings would be possible without the 
assistance of radial velocity data.
The main source of uncertainty in ring detection from an asymmetric light curve is the possibility that the asymmetry is created by an  eccentric orbit of the planet.
The fine structure of the light curve for the planet on eccentric orbit would be rather different than the structure for most of the curves for a ringed planet, as we discuss in Section \ref{sec:results_eccentric_orbit}.

The third lesson that we can draw from Fig. \ref{fig:ring_summary} is that in the case of face-on orbits (right column), both radial velocity observations and precise photometry would give no signal 
for a non-ringed planet.
A ringed planet, however, typically produces a double brightness maximum, as the rings are illuminated first from the observer's side and then from the back side during the orbit (dotted curve). 
At obliquity $\epsilon=89$\deg$ $ (solid curve), 
the rings are not visible because the observer is nearly in the ring plane.
The observer is also unable to see the rings at $\epsilon=1$\deg$ $ (dashed curve) because the star always lies (nearly) in the ring plane.
These extreme geometries are not typical for the random observations; 
more typical intermediate geometries ($\epsilon=45$\deg, dotted curve) result in a double peak, which cannot be 
confused with a light curve for a non-ringed planet on a circular orbit.
Only for rare geometries producing equal height of the two peaks, and in the absence of a detailed measurement of the curve's shape, could 
these two maximums be confused with a curve for a 
planet with half the orbital period.
Generally, the maximums will differ in height, indicating that both maxima belong to a single orbit.
Double maxima can be produced in eccentric orbits, but 
these exhibit light curves with differing fine structure (see Section \ref{sec:results_eccentric_orbit}).
We note, however, that a double peak may be generated by 
seasonal variations or by an uneven brightness distribution 
at the planet's surface. For example, a planet with bright poles rotating ``on its side'' (rotation axis parallel to ecliptic) 
whose orbit is observed face-on would exhibit two maxima in its 
light curve.

\subsection{Effects due to eccentric orbits}
\label{sec:results_eccentric_orbit}

\begin{figure}[htbp]
\plotone{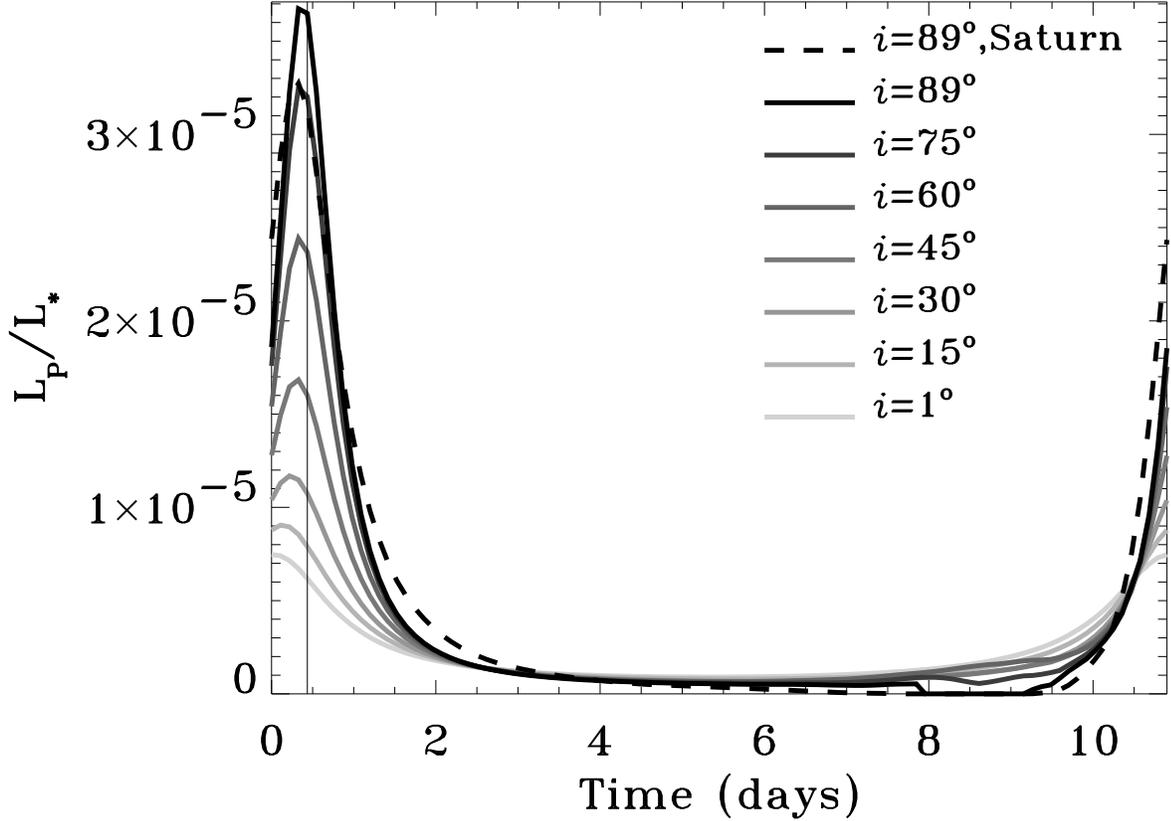}
\vspace{-1.8in}
\caption{Light curves for the planet HD~108147b, 
assuming Jupiter's equatorial radius and surface 
scattering properties, and 10\% oblateness.
The argument of pericentre is $\omega=-41$\deg.
The corresponding time of the maximum phase $\Theta=0$\deg$ $,  shown by the vertical line, is $\sim$ 0.4 days after pericentre (which defines time zero).
Different orbital inclinations from face-on to edge-on orientations ($i$=1\deg$ $ to $i$=89\deg$ $, respectively) 
are shown by lines of varying greyness. 
An edge-on light curve for a planet with Jupiter's radius but  Saturn's surface properties is shown as a dashed line.} 
\label{fig:real_planet}
\end{figure}

\corc{One of the primary differences between extrasolar 
planets discovered to date and those in the Solar System 
is the large range of orbital eccentricity displayed by 
exoplanets.  As an example of the effect eccentricity may 
have on the light curve of a planet in reflected light, 
\cora{Figure \ref{fig:real_planet} shows light curves for 
exoplanet HD~108147b, assuming the equatorial radius and atmospheric properties of Jupiter.}}

Prescribing Jupiter's scattering properties to such a close-in planet is a questionable assumption.
HD~108147b must be much hotter than Jupiter and may not have clouds at all or may have dark silicate or iron clouds \citep{sudarsky03}.
Nevertheless, if clouds are present on this planet, the scattering of properties of Jupiter may roughly approximate the phase function and yield an order of magnitude estimate of the planet's brightness (or, more correctly, an upper limit to the 
brightness, since Jupiter's clouds are nearly white).

In Fig. \ref{fig:real_planet}, we assume the planet 
to be 10\% oblate, as is Saturn. 
\cora{The planet's luminosity $L_P$ is normalized by the star's luminosity $L_*$, and is plotted versus time.}
We plot the light curve over one complete orbit, beginning at  pericentre.
The orbital parameters are those determined for HD~108147b 
by precise radial velocity measurements \citep{Pepe02}.  We 
have chosen to model HD~108147b because of its high eccentricity ($e=0.498$) and relatively short semi-major axis ($a=0.104 $AU), which accentuates the effects of orbital eccentricity. 
Since the inclination of the orbital plane is not determinable from radial velocity measurements, we display light curves for  different inclinations between $i$=90$^{\circ}$ (edge-on) and $i$=0$^{\circ}$ (face-on).
\cora{The azimuthal location of the pericentre relative to the observer can be derived from the radial velocity measurements.}
\cora{For this HD~108147b, the argument of pericentre is $\omega=-41$\deg, which means that we are fortunate to be at the azimuth at which the fullest planet phase $\Theta=0$\deg$ $ is separated from pericentre by only  41\deg.
With such a geometry, the phase-induced and eccentricity-induced maxima on the light curve amplify one another.
As a result, the amplitude of the curve for the edge-on case is about 3.5\tms10$^{-5}$, nearly five times larger than in the face-on case, where only orbital distance variation matters.
Because planets with edge-on orbital inclinations are detected 
more easily via radial velocity measurements, the chances of seeing an edge-on planet with a large light curve amplitude in follow-up observations is rather high.}

The light curves of Fig. \ref{fig:real_planet} can be rescaled 
easily for a planet with the same orbital eccentricity but  different semi-major axis $a_1$ by multiplying the luminosity by $(a/a_1)^2$ and multiplying the time axis by the ratio of the orbital periods $(a_1/a)^{3/2}$.

\begin{figure}[htbp]
\plotone{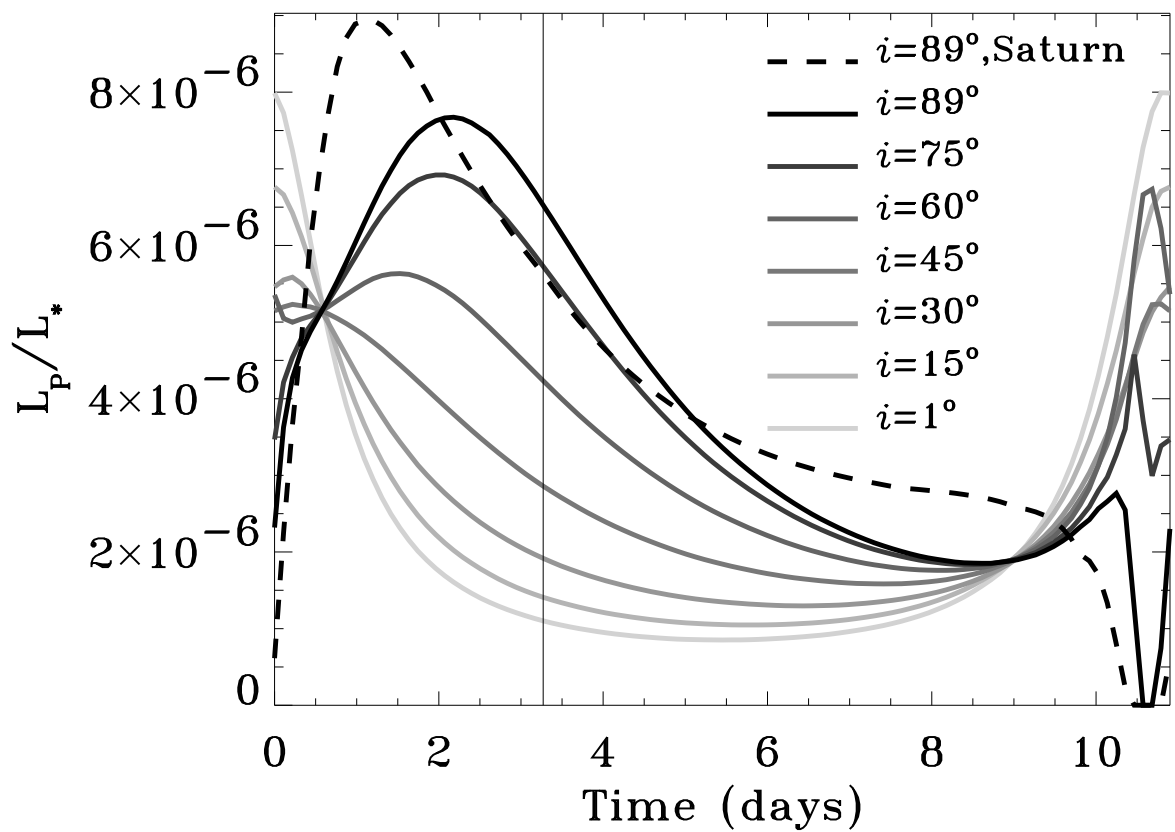}
\vspace{-1.8in}
\caption{\corc{Light curves for a planet with the orbit 
of HD~108147b, but with an argument of pericentre 
$\omega=60$\deg.
The planet has the radius and surface scattering 
properties of Jupiter.
The time corresponding to maximum phase is shown by the vertical line, and is $\sim$ 3.3 days after the pericentre.
The orbital inclination changes from face-on to edge-on orientations ($i$=1\deg$ $ to $i$=89\deg$ $ respectively), 
as the line color changes from light grey to black. 
An edge-on light curve for same planet, but 
with Saturn's surface scattering is shown as a dashed line.}
}
\label{fig:real_planet_b30}
\end{figure}

\corc{In order to illustrate the importance of the argument 
of pericentre in determining the light curves of a 
given system measured by the observer, Figure \ref{fig:real_planet_b30} shows model light curves for planet 
HD~108147b with the same parameters in assumed for 
Fig. \ref{fig:real_planet}, 
except that the argument of pericentre is now 
set to $\omega=60$\deg, rather than the true $\omega=319$\deg.} 
\cora{This example also serves to demonstrate the key features of the light curves for eccentric orbits.}

Light curves for orbits observed nearly 
edge-on ($i\approx 90^{\circ}$) show a primary peak when the planet  presents its full phase to the observer.  
The amplitude of the reflected light at this peak is approximately $L_{P}/L_{*}$=6.5 \tms 10$^{-6}$.  
The amplitude of this primary peak is reduced considerably 
at lower inclinations, since the planet no longer exhibits full phase at the maximum phase position.  
Additionally, the position of this peak moves towards the pericentre as the inclination is lowered.

The edge-on light curve also displays a secondary peak located close to pericentre.  This peak is due to the increase 
in reflected light corresponding to the smaller planet-star 
separation at pericentre.
At lower inclinations, this secondary peak becomes the higher 
of the two peaks as the planet-star separation function dominates  over the phase function.  
In the extreme case of $i=0^{\circ}$, the phase function no longer affects the light curve (the planet displays a constant half phase), and the light curve is purely a function of the planet-star separation.

The $i \approx 90^{\circ}$ and $i=75^{\circ}$ light curves are complicated by the appearance of a sharp trough where the amplitude of reflected light is reduced almost to zero.  This trough is due to the planet showing no phase (``new moon'') 
at this point of its orbit, and is particularly sharp due to the planet's high angular speed at this point in its orbit.  The trough is not prominent in lower inclinations because the minimum phase of the planet is no longer close to zero.

\cora{The double peaks are rather rare; 
for most values of $\omega$ they do not appear at any orbital inclination.
If the planet's surface is not as forward-scattering as 
is Jupiter's (\eg, Saturn), the second peak is even less likely to appear.
However, when the second peak is present, it is necessarily accompanied by the sharp trough immediately next to it.
This feature makes the double-peaked light curve of an eccentric orbit distinguishable from the double-peaked light curve of a  ringed planet, for which the two peaks are broad (see Section \ref{sec:ring_effects}).}

\cora{Single-peaked curves from eccentric orbits or ringed planets  may be distinguished even without radial velocity data because the curve is smooth in the eccentric case.
Although both ringed planets and eccentric orbits may result in an asymmetric light curve, the complicated shadowing of a ringed planet creates abrupt variations along the curve (see Fig. \ref{fig:ring_summary}).}

\corb{The location of the light curve maximum relative to the pericentre can yield important constraints on the geometry 
of the orbit and the cloud cover of the planet.}
\cora{First, the orbital inclination is restricted by 
the amount of the temporal shift in the light curve maximum from pericentre.
As the inclination increases, the maximum moves from 
pericentre (time zero in Figs. \ref{fig:real_planet} and \ref{fig:real_planet_b30}) toward the time of the maximum phase  $\Theta=0$\deg (the vertical line in Figs. \ref{fig:real_planet} and \ref{fig:real_planet_b30}).
This shift is larger when the orbit is viewed from a position that places pericentre close to half-phase ($\omega\sim$ 0\deg or $\omega\sim$ 180\deg) rather than ``new moon'' ($\omega\sim 90$\deg) or ``full-moon'' ($\omega\sim -90$\deg) phases.
Unlike the amplitude of the curve, which is also an indicator of inclination, the shift cannot be alternatively produced by 
altering the planet's size or albedo.
However the shift can result from a different strength of backscattering at the planet's surface. 
An example of such ambiguity is the edge-on light curve 
for a planet with Saturn's scattering properties 
(dashed line in Fig. \ref{fig:real_planet_b30}), which has a maximum where a Jupiter's curve of $i\approx$50\deg$ $ would have its maximum.
The ambiguity may be resolved by comparing the detailed shape of the curves.
}

\cora{Second, the curve's maximum shift may also serve to restrict the surface properties of the planet.
Although the shift can be confused with the effect of orbital inclination, a lower limit may be put on the strength of backward scattering for the planet's surface. 
\corb{For each type of scattering surface}, a maximum possible shift \corb{of the curve's peak from pericentre towards $\Theta=0$\deg} occurs in the face-on case \corc{($i=0$\deg)}.
If a larger shift is observed, it can only be explained by a sharper backward scattering peak
of the surface.
Constraining surface properties in this way 
should be treated with caution, however, because of possible 
contamination by rings (we did not model both rings and eccentric orbits simultaneously), large-scale bright patches on the planet's surface, or seasonal variation of the cloud coverage.
}

\begin{figure}[htbp]
\plotone{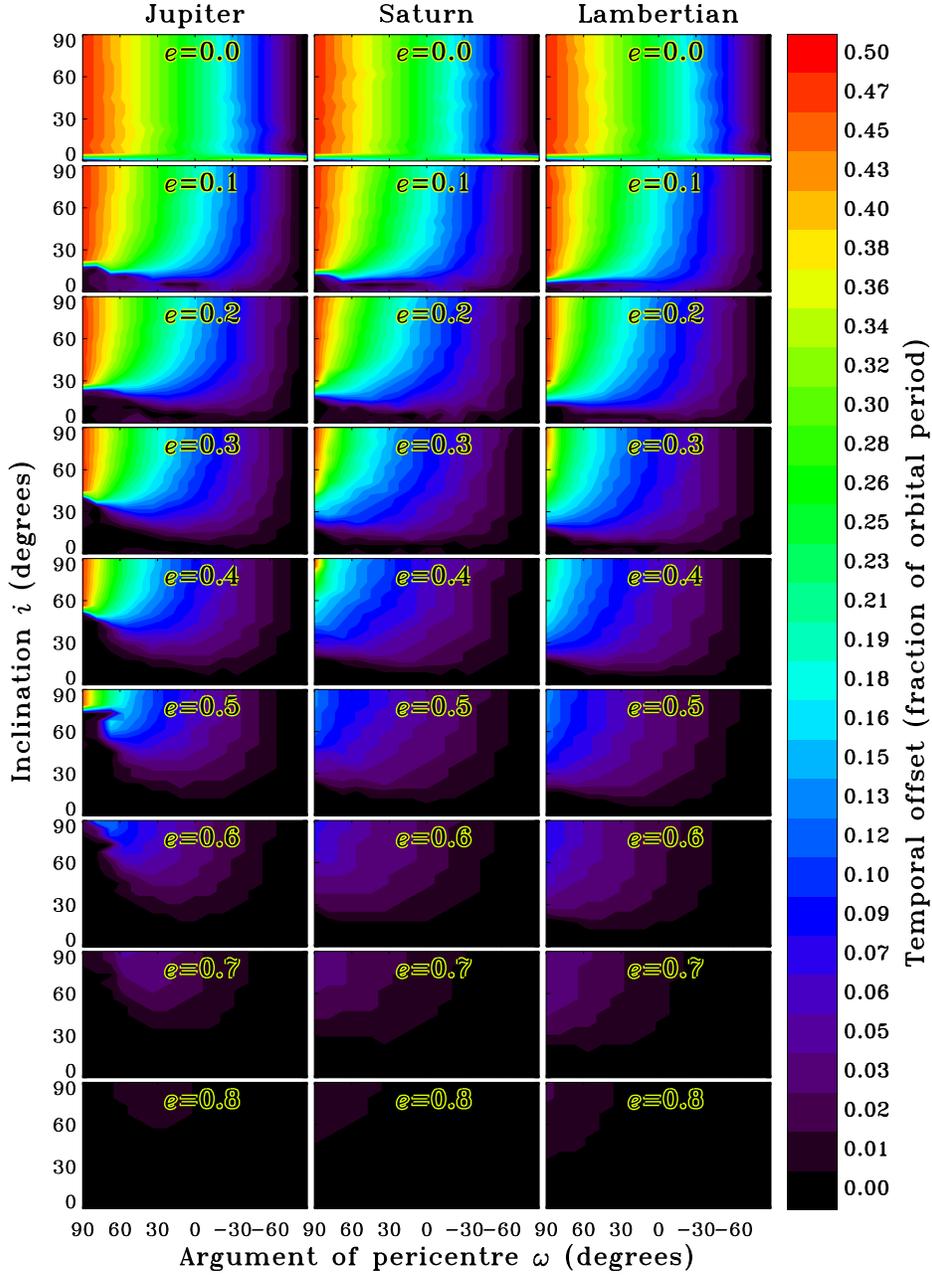}
\vspace{-1in}
\caption{
\corc{Summary of the temporal shift of the light curve maximum from the pericentre for eccentric orbits at different observational geometries for planets with Jupiter-like (left column), Saturn-like (middle column), and Lambertian scattering properties.  Shifts are measured as a fraction of 
the orbital period.  
Each planet is assumed to be 10\% oblate.
The eccentricity of the orbit increases from the top 
row of subplots to the bottom row. Viewing geometry is 
described by the orbital inclination and argument of 
pericentre, plotted on the ordinate and abscissa, respectively 
of each subplot.}}
\label{fig:dt_maps}
\end{figure}

Figure \ref{fig:dt_maps} illustrates the sensitivity of the 
shift 
to the inclination 
of the orbit (displayed along the ordinate of each subplot) 
and different atmospheric scattering (each of the 
three columns of subplots).
The right side of each subplot ($\omega=-90$\deg) corresponds to observations from the apocentre side of the orbit.
At such a position, the planet is observed at its fullest possible phase ($\Theta=0$\deg) at pericentre, and 
since the eccentricity-induced and phase-induced maxima in the light curve coincide at pericentre, the shift between the light curve maximum and pericentre is zero. 
The left side of each subplot ($\omega=90$\deg) corresponds to observations from the pericentre side of the orbit.
Here the planet is observed at its smallest possible phase ($\Theta=180$\deg) at pericentre, so that the eccentricity-induced maximum is shifted from the phase-induced maximum by half the orbital period.
The maximum of the combined light curve then has a complicated dependence on $e$, $i$ and planet's scattering properties.
In the simple case of $e=0$, the eccentricity-induced maximum does not exist, and the curve's maximum is always at the fullest phase. 
In this case, the time of the maximum follows the observer's position $\omega$, with the shift reaching half phase at $\omega=90$\deg$ $ (red at the left on all $e=0$ plots).
 
If radial velocity data exist for the planet, then 
the orbital eccentricity $e$ and argument of pericentre $\omega$ are known, and the shift, indicated by the color coding 
in Fig. \ref{fig:dt_maps} is observable.
The orbital inclination $i$ (shown on 
the ordinate) is generally not known.
\corc{Orbital inclination is known for transiting 
planets ($i\approx 90$\deg), but is only very 
loosely constrained in the absence of a transit 
to $i<(90-\Delta)$\deg, where $\Delta$ is a small angle depending on the planet's size and planetary system's geometry.}

Our method of constraining the orbital inclination $i$ 
is achieved by matching a measured time shift value 
with those along the vertical line corresponding to the known $\omega$ on the plot with the appropriate eccentricity, 
as measured by radial velocity techniques.
For example, a planet with Jupiter's scattering 
properties on an eccentric $e$=0.4 orbit, observed to 
have a temporal shift of 20\% of the orbital period (green) by 
an observer with $\omega \approx$45\deg$ $, 
would imply inclinations of 60--90\deg.
A shift of 10--15\% of the orbital period (light blue) 
would mean orbital inclinations of 30--60\deg, while shifts 
smaller than 10\% (purple to black), would mean inclinations of 0--30\deg.
An ambiguity remains due to the unknown scattering properties 
of the planet's surface.
Three examples of the scattering are shown in the three columns 
of Fig.~\ref{fig:dt_maps}: a strongly backward scattering Jupiter, a weaker backward scattering Saturn, and the less realistic extreme case of Lambertian surface.
Back scattering stronger than that exhibited by Jupiter may exist on extrasolar planets. We did not model this case, but 
expect that such an extreme example would produce 
a column of plots to the left of the Jupiter column, 
following the trends back scattering increases from Lambertian on the right to Jupiter on the left.

\corc{Examination of the columns for Jupiter, Saturn and Lambertian surfaces indicates that in some cases, scattering properties 
may be inferred from observation of the temporal 
shift in the reflected light.
In some geometries a strong temporal shift may be observed at given $e$ and $\omega$.
For example, if $e=0.4$, $\omega$=60\deg$ $, and the 
observed shift 
is 25\%  the orbital period, only the left plot corresponding to Jupiter-like 
scattering displays green anywhere along the $\omega$=60\deg$ $ vertical line; thus only strongly back scattering planet like Jupiter can be consistent with the observation.}

The contrast, 
or difference between the maximum and minimum amplitude of the light curve, gives a measure of the degree of variability that the planet's reflected light displays.  It is this variation that new generation space or ground-based photometers 
may be able to detect if they can achieve the 
required levels of precision.  The largest contrast is 
observed for edge-on, $i=90^{\circ}$, orbits \cora{(see Fig. \ref{fig:real_planet_b30}}), while at lower inclinations the contrast is reduced.  However at very low inclinations approaching $i$=0$^{\circ}$, the contrast increases again as the amplitude of reflected light at the pericentre increases.  Since the contrast at $i$=0$^{\circ}$ is higher than at mid inclinations, 
measuring reflected light may provide a method of detecting planets with face-on orbits, so long as the planet's orbit is highly eccentric.  Reflected light observations 
would then complement other methods, such as radial velocity measurements, that are unable to detect planets orbiting with face-on orientations.

\begin{figure}[htbp]
\plotone{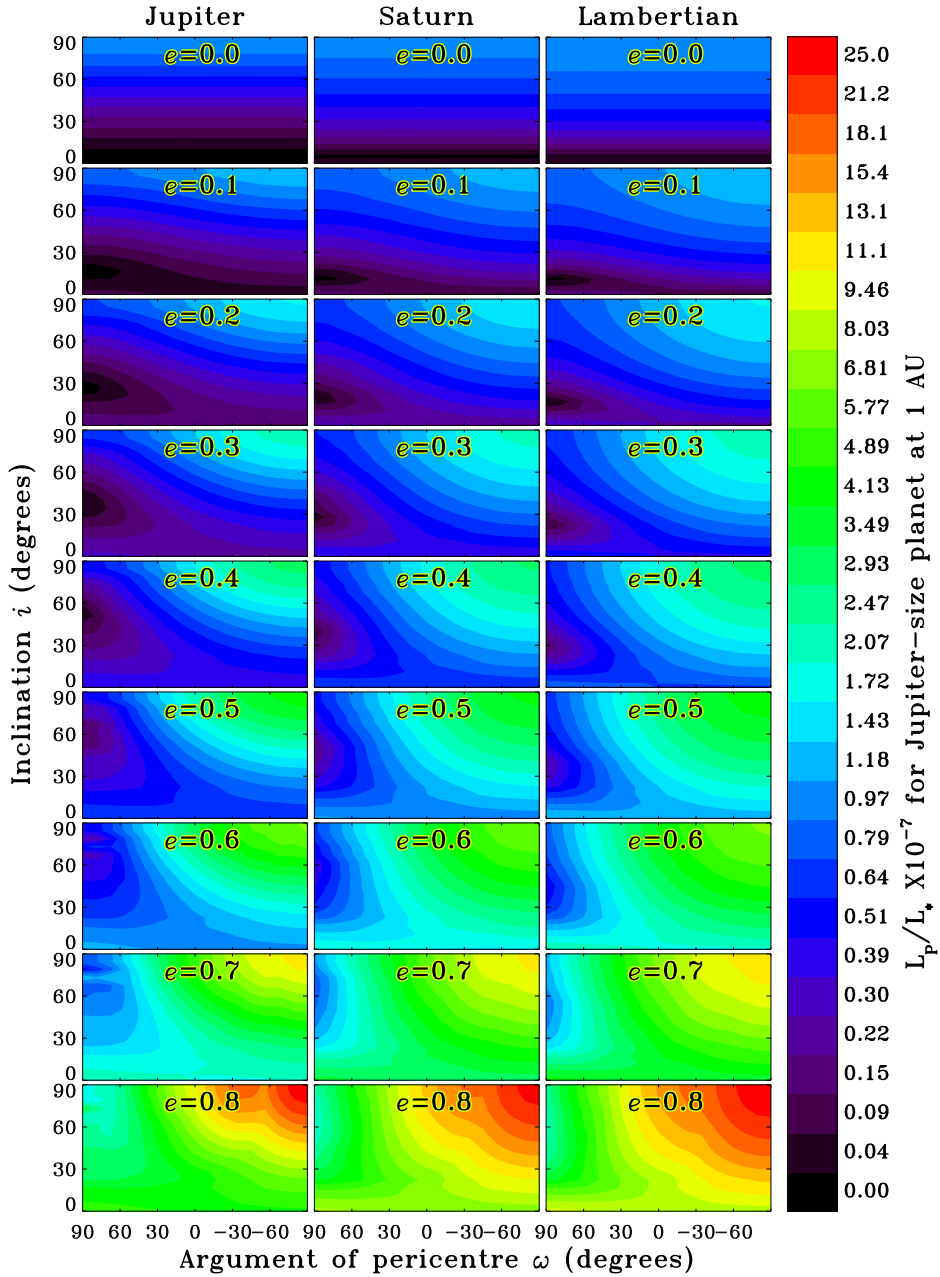}
\caption{
Summary of light curve variability amplitudes, 
or contrast, for eccentric orbits at different observational geometries for Jupiter (left column), Saturn (middle column), and Lambertian scattering properties 
with surface albedo 0.82 (right column).
Planets are assumed to be 10\% oblate.}
\label{fig:observability}
\end{figure}

We plot the light curve contrast for all possible orbital orientations in Figs. \ref{fig:observability}.  
Note that the backward scattering peak of Jupiter's surface makes the planet much darker at low inclinations than Saturn or Lambertian planet. 
These contour maps display the degree of contrast for various orbital inclinations and arguments of pericentre at given eccentricities.  At a given orientation, the more eccentric orbits show much higher contrast than circular orbits, although 
the amount of contrast is 
strongly dependent both on inclination and argument of pericentre.  At high eccentricities, favorable geometries (such as  $i=90^{\circ}$ and $\omega=-90$) can increase the contrast by approximately five times that of less favorable orientations.


\section{Discussion}
\label{sec:discussion}

\subsection{Uncertainties}
\label{sec:uncertainties}
Our model provides a rather accurate description of 
reflected light curves for Jupiter and Saturn.
The largest uncertainties are due to the lack of observations at 
large phase angles (forward scattering at $\alpha>$150\deg$ $) for the surfaces of Jupiter, Saturn, and its rings. 
The extrapolations we have made 
to the phase functions could result in a factor of a few 
error in the brightness at these angles.
Luckily, the luminosity of the planet at forward-scattering geometries is small because of the small area of the 
illuminated crescent at those positions.
Consequently, this uncertainty does not influence 
the large amplitude features of our modeled light curves.
\corc{The maxima of the highest-amplitude (edge-on) 
light curves, which happen at opposition, can be constrained to 
within a few percent by high spectral resolution ground-based 
data for Saturn and Jupiter.
In our study, the uncertainty of these maximum amplitudes is 10-20\%  because the brightness of a planet varies with wavelength by about the same amount within the 0.6-0.7 \mm spectral range \citep{karkoschka98}.} 

Reflectivity of the rings may be in error by as much 
as 50\% because many geometries, especially those 
for face-on ring illumination and face-on ring observation, 
are not constrained by observations.
Face-on illumination never happens for Saturn's rings because of its 27\deg$ $ obliquity.
Observations of the rings face-on are unavailable because all spacecraft visiting Saturn have had trajectories close to ecliptic. 
New observations by the Cassini spacecraft (arriving at Saturn in July 2004) will fill this gap in the data. 
Another source of the ring error is the coarse grid in the tabulated ring scattering properties (see Section \ref{sec:model}).
This grid-induced noise is usually below 30\% of the ring's luminosity.

Rings around close-in, short-period exoplanets may be unlikely 
because of tidal disruption, and in any case can only consist of rock. (Saturn's rings are 99\% ice.) 
Ice rings are stable against evaporation outside $\sim$7 AU for solar-type stars \citep{mekler94}. 
Rocky ring material would probably be several times darker 
than ice, but would remain backward scattering for geometric reasons.  \corc{Indeed, the rocks in Saturn's rings 
behave as moonlets and are bright at backward-scattering full-phase geometry.}
Ring density for extrasolar planets is difficult to predict.
It is possible that rocky rings of extrasolar planets are massive and optically thick.
An example in the Solar System is Earth, which 
may have had massive rings while the Moon was forming.
Furthermore, ring size for exoplanets may be very different from Saturn's rings, which 
would have a large effect on the ring luminosity.
We do not model various ring sizes here and refer the reader to the work of \cite{arnold04} which investigates ring size effects.
The optical depth of any rings around extrasolar planets may 
also differ substantially from that of the rings of 
Saturn. 
It is important to note that when a ring has 
an optical depth more than about 3-5, 
the brightness of illuminated side becomes insensitive to further increases of optical depth.
More than half of the area of Saturn's rings 
has an intermediate optical depth of 0.5-3, which can be considered neither optically thin nor optically thick. 
These optical depths cause 
the dark side of Saturn's ring to be quite bright.
If the rings of an extrasolar planet are denser than Saturn's, the ring-induced light curves will be more asymmetric due to the difference between the dark and bright sides of the rings.
If the rings are optically thin, some asymmetry due to shadows will remain, but the light curve asymmetry due to the different sides of the ring will decrease. 
Since most of Saturn's rings are quite dense, the brightness of the illuminated side is nearly equal to 
the ``saturated'' brightness achieved at infinite optical depth.
\corc{For this reason, we do not expect that an 
increased optical depth in any extrasolar rings will result in a large increase over what we predict here 
in the brightness at the light curve maximum.
The effects of differences in ring size or 
ring particle albedo are likely to be more prominent.}

Application of our model light curves for the bodies 
of Jupiter and Saturn to the bodies of 
close-in giant exoplanets, whose bright light curves will 
be the most observable, has a number of complications related to 
chemistry.
The clouds on these exoplanets are expected to be composed of solid Fe, MgSiO$_3$, Al$_2$O$_3$, and other condensates 
that are stable at high temperatures \citep{seager00, burrows99}, 
rather than the water and ammonia ices seen in Solar System giants \citep{weidenschilling73}.
This clouds may be darker and the corresponding light curve amplitudes may be several times smaller.
\corb{For spherical particles, the scattering phase function depends on size of the particle compared to the wavelength of light, and on particle composition.}
\corc{
Consequently, it is possible that forward and backward scattering maxima typical of Jupiter and Saturn could occur for cloud-covered exoplanets even if their cloud composition is different.}
Remarkably, the direct spectral signature of ammonia or water ice particles is found in small weather patches 
that cover only a few percents of Jupiter's area \citep{simon00,baines03}.
Most of the area is covered by clouds that 
exhibit no signature of ice on the surface of the particles.
This may be due to the coating of the condensate particles by photochemically produced materials.
Whether the photochemistry on extrasolar giant planets could alter the surface of cloud particles is beyond 
the scope of this paper, but direct comparison of clouds made of pure silicate on extrasolar planets, for example, with clouds made of pure ammonia on Jupiter and Saturn is probably oversimplified.   
The most important parameter in determining 
the brightness of a planet is the presence or absence of clouds, which depends on poorly-known (even for Jupiter and Saturn)  vertical atmospheric circulation.
Detection or non-detection of a light curve with the expected amplitudes would be most indicative of such cloud presence.
\subsection{Prospects for Detection by Modern Instruments}
\label{sec:detectability}

Short-period giant planets are the easiest targets for detection 
by precise (spatially unresolved) 
photometry because they are well-illuminated by the star.  
As discussed in the last section, light curves for such planets may differ from the 
those we present here for Jupiter and Saturn because their hotter 
temperatures (and possibly different evolution as a result 
of migration) will alter the planet's atmosphere and clouds and thus its scattering properties.

Nevertheless, our model light curves do provide 
a first-order approximation for a cloud-covered exoplanet 
and so we will discuss the model light curves of 
Jupiter and Saturn as if they were extrasolar planets placed at small orbital distances.
In that case, our 
light curve amplitudes (Figs. \ref{fig:jup_sat} to \ref{fig:ring_summary}) can be converted to measurable units 
of fractional luminosity $L_P/L_*$ using Eq. \ref{eq:lp_l*}.
The ratio $L_P/L_*$ depends inversely on the square of the planet's 
orbital distance, and is independent of the star's distance 
to the observer.
Typical extrasolar planets discovered to date have orbital 
distances (semi-major axes) between 0.3 and 3 AU, 
corresponding to conversion factors of 2.5\tms10$^{-4}$ to 2.5\tms10$^{-8}$ multiplying the order-of-unity light curve amplitudes displayed in Figs. \ref{fig:jup_sat} to \ref{fig:ring_summary}, assuming that the planet 
has a radius equal to that of Jupiter's at 71400 km. 
For convenience, we have scaled amplitudes in 
Fig. \ref{fig:observability} to a 
planet with a semi-major axis equal to 1 AU.

The ability of precise space-based photometers such as 
MOST and KEPLER to detect reflected light from extrasolar planets depends on the signal-to-noise ($S/N$) of the observations, 
and thus the apparent brightness of the target parent 
star.  Precisions approaching $10^{-6}$ are expected for 
stars as bright as 6th magnitude or brighter in V band ($V \leq 6$) 
with MOST \citep{green03,walker03}, 
and $\sim67 \times 10^{-6}$ for stars as faint as $R = 12$ in $R$-band\footnote{\corc{We model wavelengths here that are more similar to 
the $R$ passband than the $V$ band.}} with KEPLER 
\citep{jenkins03,koch00,jenkins00,remund01} independent of 
magnitude.
Within a fiducial survey volume of radius 30pc, 
where a solar-type star would have $V \leq 7.2$ and $R \leq 6.9$, 
MOST and KEPLER would thus be able to provide measurements with 
precisions of $\sim2 \times 10^{-6}$ and $\sim6 \times 10^{-6}$, respectively, except where they are dominated 
by instrumental or stellar noise, which may 
\corb{decrease} 
the sensitivity by factors of five for the faintest objects.
For short-period planets, the $S/N$ may be increased by repeated 
observations over several orbits.  
On the other hand, 
\corb{planets with periods longer than the order of several days}
may suffer more from confusion with fluctuations 
in stellar luminosity on the order of $10^{-5}$ \corb{\citep{jenkins03}}. 
In summary, space-based photometers currently launched such 
as MOST (with no survey capability) and about to be launched (KEPLER, 
a survey instrument), should be able to detect variations in 
the total light received from stars within 30pc due to reflected light 
from orbiting planets at the level of $10^{-5}$. 

\corc{Direct, spatially-resolved imaging, of the sort that 
may be provided by future space instruments or 
adaptive optics on Extremely Large Telescopes (ELT) 
from the ground (both aided with coronagraphic  
or nulling techniques) may detect 
faint planets if they are sufficiently distant from the star 
to be separable from the angular size of the optical 
point spread function 
\citep{dekany04,lardiere03,codona04,trauger03,krist03,clampin01}.
}

\corc{For example, \cite{lardiere03} used detailed simulations 
of ideal adaptive optics systems with different actuator pitches 
placed on ground-based ELT of various aperture sizes located 
in varying atmospheric conditions, to calculate the 
planet-to-star flux ratio that would be required 
in order to reach 
a signal-to-noise of three ($S/N = 3$) in one night's 
observation (10 hour exposure).  The planet-star flux ratio 
is related to, but could be slightly larger than the 
contrast values of Fig. \ref{fig:observability}, which 
show only the {\it varying\/} portion of the planet's 
brightness compared to the (assumed) constant stellar 
brightness. 
Generally speaking, they conclude that a 30m-diameter ELT 
on one of the best sites in the world (\eg\ Mauna Kea or in the 
Antarctic) could easily detect with $S/N=3$ planets with 
brightness ratios as small as $1 \times 10^{-9}$ over 
the range of projected planet-host star separations 
corresponding to $0.1 - 1.0 $ 
arcsec.  These separations correspond to planets on orbits with
semi-major axes from 1 to 10~AU for systems 10pc distant 
from Earth (and about a factor of ten greater at 30pc). 
Such planets are more likely to have the temperatures 
of Saturn and Jupiter, and thus the scattering 
properties we have assumed here.  
We note that a ratios of $1 \times 10^{-9}$ is about 100 times 
the bright, dark blue contrast levels shown in 
Fig. \ref{fig:observability}, 
indicating that most systems we have considered would 
indeed be detectable even if they orbited 10~AU from 
their host.}

\corc{Projected onto the sky plane, a planet is most separated 
from its host star near half-phase, \ie, at $\Theta\sim\pm 90$\deg\ (Figs. \ref{fig:jup_sat} to \ref{fig:ring_summary}).
At these points the planet is not at its maximum brightness, 
but may still be detectable.  Note too that Lambertian planets may underestimate amplitudes at half-phase 
of planets that scatter anisotropically (see Fig. \ref{fig:jup_sat}).
On the other hand, rings may increase the luminosity by a factor of $2-3$ or more if the rings are larger than those of Saturn, 
increasing the chance of light curve detection, as would 
larger planets.
Many technical challenges will have to be overcome 
before contrast levels of $1 \times 10^{-9}$ become 
detectable.  
In this light, it is interesting to note that because planet size is 
not believed to grow substantially with planet mass above 
one Jupiter mass \citep{guillot96}, searching for the 
photometric signature of large rings of 
extrasolar planets (if they exist) 
may be an easier task than searching for the planets themselves.}


\newpage
\section{Conclusions}
\label{sec:conclusions}

\corc{We have examined the effects of rings, realistic scattering properties 
(\ie\ those actually measured for Saturn and Jupiter), 
viewing geometry, and orbit eccentricity on the characteristics of 
reflected light from extrasolar planets, and their combined planet-host 
star light curve.
In particular, we have noted signatures of ringed planets, and 
have indicated cases in which planetary rings may be distinguished 
from, or confused with, other effects.}

\corc{In the following table, we summarize signatures of planetary rings on 
the light curves of their extrasolar planets, and note differences 
between ring effects and those of a non-ringed planet on an eccentric 
orbit, or planetary surface effects.}

\begin{table}[htbp]
\begin{small}
\corc{
\begin{tabular}{|l|lll|}
\hline
Signature   & Rings    &Eccentric orbit                               &Alternative\\
\hline
Light curve amplitude   & increases             & increase               & larger albedo \\
                                & by factors $2-3$\tablenotemark{a} & depends on $e$ & or planet size\\
\hline
Double peak             & rare                           & rare, sharp trough & surface asymmetry\tablenotemark{b} \\
                        & smooth peaks     & near one peak      & or seasons\\
\hline
Large-scale asymmetry   & common               & common               & surface asymmetry\tablenotemark{b},\\
                        &                      &                      & oblateness, or seasons\\
\hline
Fine light curve structure&abrupt changes &smooth                                  &  \\
\hline
Shift of curve max      & often                & never for circular  & surface asymmetry\tablenotemark{b},\\
from $\Theta$=0\deg     &                      & orbit,calculable    & oblateness, seasons, or \\
(with radial velocity data) &                  & for eccentric orbits& back scattering\tablenotemark{c}\\
\hline
\end{tabular}
}
\end{small}

\corc{\tablenotetext{a}{For Saturn-sized (relative to the planetary radius) rings.}
\tablenotetext{b}{Planet-scale asymmetry of the brightness distribution, \eg, bright poles.}
\tablenotetext{c}{The strengh of the back-scattering peak of the planetary phase function changes the shift 
of the curve maximum relative to $\Theta=0$\deg, which is known within the range defined by the possible phase functions (see Fig. \ref{fig:dt_maps}).}
}
\end{table}

\corc{Our studies of planets with Jupiter-like, Saturn-like and Lambertian scattering 
properties on eccentric orbits hold the following messages for potential observers:
\begin{enumerate}
\item An anisotropically scattering planet is much fainter at half-phase ($\Theta$=$\pm$ 90\deg) than is a Lambertian planet.
\item Anisotropically-scattering planets are also much fainter at low inclinations than are Lambertian planets.
\item For many geometries, eccentricity of the orbit may increase light curve amplitude by a large amount compared to a planet on a circular orbit with the same semi-major axis (see Fig. \ref{fig:observability})
\item The timing of the lightcurve maximum with respect to pericentre may be 
used to restrict the orbital inclination and atmospheric back scattering properties 
for a large fraction of possible geometries (see Fig. \ref{fig:dt_maps})
\end{enumerate}
In summary, rings, eccentric orbits, and atmospheric scattering 
properties of exoplanets may be detected with future telescopes in 
the next decade or so by the effect they have on the light curve of the 
planet's reflected light, and these effects may often be distinguished from one another 
by the shape of the observed light curve alone or with the aid of 
additional radial velocity data.
}
 

\section*{Acknowledgements}
We thank R.A. West for useful references on Jupiter's scattering.
The work begun at NASA GISS while U.D. was supported by Anthony D. Del Genio under the Cassini Project.
\bibliography{my}
\label{'last page'}

\end{document}